\documentclass[a4paper]{report}
\usepackage[utf8]{inputenc}
\usepackage[T1]{fontenc}
\usepackage{RJournal}
\usepackage{amsmath,amssymb, array}
\usepackage{booktabs}
\usepackage{multirow}
\usepackage{graphicx}
\usepackage[ruled,vlined]{algorithm2e}

\usepackage{subfigure}

\newtheorem{Definition}{Definition}
\newcommand\independent{\protect\mathpalette{\protect\independenT}{\perp}}
\def\independenT#1#2{\mathrel{\rlap{$#1#2$}\mkern2mu{#1#2}}}
\begin{document}
 
\sectionhead{Contributed research article}
\volume{XX}
\volnumber{YY}
\year{20ZZ}
\month{AAAA}
\newtheorem{Theorem}{Theorem}[section]
\newtheorem{Lemma}{Lemma}
\def\argmin{\mathop{\rm arg\, min}}

\begin{article}
\title{itdr: An R package of Integral Transformation Methods to Estimate the SDR Subspaces in Regression}
\author{by Tharindu P. De Alwis, S. Yaser Samadi, and Jiaying Weng}

\maketitle

\abstract{
Sufficient dimension reduction (SDR) is an effective tool for regression models, offering a viable approach to address and analyze the nonlinear nature of regression problems. This paper introduces the \textbf{itdr} R package, a comprehensive and user-friendly tool that introduces several functions based on integral transformation methods for estimating SDR subspaces. 
In particular, the \textbf{itdr} package incorporates two key methods, namely the Fourier method (FM) and the convolution method (CM). These methods allow for estimating the SDR subspaces, namely the central mean subspace (CMS) and the central subspace (CS), in cases where the response is univariate. 
Furthermore, the \textbf{itdr} package facilitates the recovery of the CMS through the iterative Hessian transformation (IHT) method for univariate responses.
Additionally, it enables the recovery of the CS by employing various Fourier transformation strategies, such as the inverse dimension reduction method, the minimum discrepancy approach using Fourier transformation, and the Fourier transform sparse inverse regression approach, specifically designed for cases with multivariate responses.
To demonstrate its capabilities, the \textbf{itdr} package is applied to five different datasets. 
 Furthermore, this package is the pioneering implementation of integral transformation methods for estimating SDR subspaces,  thus promising significant advancements in SDR research. 
 }
 
\section{Introduction: Sufficient Dimension Reduction in Regression}
Let $Y$ be a univariate response, and let $\mathbf{X}$ be a $p$-dimensional vector consisting of continuous predictors, denoted as $\mathbf{X}=({X}_1,\cdots,{X}_p)^T$. The conditional distribution of $Y$ given $\mathbf{X}$ is denoted as $F_{Y\vert\mathbf{X}}$, and $E[Y\vert\mathbf{X}]$ represents the mean response at $\mathbf{X}$.  When $F_{Y\vert\mathbf{X}}$ or $E[Y\vert\mathbf{X}]$ lacks a specific parametric form, nonparametric methods should be employed. However, classical methods, such as polynomial smoothing methods, become impractical as the dimension of $\mathbf{X}$ increases. To address the curse of dimensionality, several dimension reduction methods have been proposed. Sufficient dimension reduction (SDR) stands out as one of the most important and successful approaches that have garnered considerable interest in recent years. The SDR method seeks to project $\mathbf{X}$ onto a lower-dimensional subspace in a manner that enables the formation of the regression of $Y$ on $\mathbf{X}$ without loss of information about $F_{Y\vert\mathbf{X}}$ or $E[Y\vert\mathbf{X}]$, thus mitigating the curse of dimensionality.

The SDR theory has its origins in the seminal work of \cite{Li91} and \cite{CW91}.  The primary objective of SDR is to estimate a $p \times d$ matrix $\boldsymbol\Gamma$, $d\leq p$, with the aim of replacing the $p$-dimensional predictor vector $\textbf{X}$ with a $d$-dimensional vector $\boldsymbol{\Gamma}^T\textbf{X}$. This substitution strives to preserve all relevant information about the responses $Y$ without any loss.  
Importantly, the SDR technique does not impose any specific model assumptions, as discussed by \cite{cook98}. In a more general context, we consider the following regression model 
\begin{equation}\label{eq:1}	Y=g(\boldsymbol{\Gamma}^T\textbf{X})+\varepsilon,
\end{equation} 
where $g(\cdot)$ is an unknown smooth link function and $\varepsilon$ denotes the error term. Within this framework, we can define the following SDR subspaces. 
\textcolor{black}{\begin{Definition}\label{def:1}
	Suppose $\mathcal{S} \subseteq \mathbb{R}^p$ with projection $\mathcal{P_S}$, and let $\mathcal{P_S}: \mathbb{R}^p\to \mathcal{S}$ be the orthogonal projection operator onto a $d~(<p)$-dimensional subspace $\mathcal{S}$.  The subspace $\mathcal{S}$ is a sufficient dimension reduction subspace if
	\begin{equation}\label{eq:2}
	Y\independent \textbf{X} \vert \mathcal{P_S}\textbf{X},
	\end{equation}
	where $\independent$ indicates independence. The intersection of all dimension reduction subspaces that satisfies Equation \eqref{eq:2} is called the central dimension reduction subspace, or \textit{\text{``Central Subspace (CS)''}}. It is denoted as $\mathcal{S}_{Y\vert\textbf{X}}$.
\end{Definition}}
The dimension of $\mathcal{S}_{Y\vert\textbf{X}}$, denoted as $d=dim(\mathcal{S}_{Y\vert\textbf{X}})$ is referred to as the structural dimension of the regression $Y$ on $\mathbf{X}$. \cite{cook98} demonstrated the existence and uniqueness of $\mathcal{S}_{Y\vert\textbf{X}}$ under certain mild conditions. Similarly, Definition \ref{def:1} can be extended to include cases where the conditional mean function $E[Y|\textbf{X}]$ is of interest. This extension leads to the concept of central mean subspace, introduced by \cite{CL02}. 
\textcolor{black}{\begin{Definition}\label{def:2}
	Suppose an orthogonal projection $\mathcal{P_S}: \mathbb{R}^p\to \mathcal{S}$ denote the projection operator onto a $d~(<p)$-dimensional subspace $\mathcal{S}$. Then, $\mathcal{S}$ is a mean dimension reduction subspace for conditional mean if
	\begin{equation}\label{eq:3}
	Y \independent E[Y\vert \textbf{X}] \vert \mathcal{P_S}\textbf{X}.
	\end{equation}
	The intersection of all mean dimension reduction subspaces that satisfies the condition given in Equation \eqref{eq:3} is called the mean dimension reduction subspace, or  ``Central Mean Subspace (CMS)''. It is denoted by $\mathcal{S}_{E[Y\vert \textbf{X}]}$.
\end{Definition}}

Several parametric and nonparametric estimation methods have been developed to estimate the CS and CMS. These include sliced inverse regression \cite[SIR;][]{Li91}, sliced average variance estimation \cite[SAVE;][]{CW91}, inverse regression estimation method \cite[IRE;][]{CN05}, and Fourier transformation method for inverse dimension reduction \citep{Weng18,weng2022minimum, weng2022fourier}, among others.  
Nonlinear dimension reduction techniques, such as kernel PCA (KPCA), generalized SIR (GSIR), and generalized SAVE (GSAVE) methods,  have been proposed to estimate the nonlinear central subspace in order to overcome the limitations of linear combinations of covariates in principal component analysis (PCA), SIR, and SAVE methods \cite[see][]{lib2018}. Moreover, \cite{eff2013} and \cite{psave2013} proposed the efficient estimator (Eff) and partial SAVE (pSAVE) methods, respectively, for estimating the CS.  Likelihood-based SDR methods, such as covariance reduction \cite[CORE;][]{cookfroz08a}, likelihood acquired directions \cite[LAD;][]{lbase}, and principal fitted components \cite[PFC;][]{cook07, cookfro08b} have been proposed for the CS estimation.
Various approaches have been proposed to estimate CMS, such as the principal Hessian direction \cite[PHD;][]{Li92}, iterative Hessian transformation \cite[IHT;][]{CL02}, and the structure adaptive method \cite[SAM;][]{H01}. The minimum average estimation \cite[MAVE;][]{X02} and outer product gradient \cite[OPG;][]{X02} methods are implemented for estimating both the CS and the CMS. Furthermore, the Fourier transformation method \cite[FM;][]{zhu2006} and the convolution transformation method \cite[CM;][]{zeng2010} have been employed  to estimate both the CS and the CMS. 

The R package \textbf{ldr}, developed by \cite{ldr2014}, implements three likelihood-based SDR methods, i.e., CORE, LAD, and PFC for estimating the CS. The R package $\mathbf{dr}$, developed by \cite{Weis2015}, implements several SDR approaches, including the SIR, SAVE, and IRE methods to estimate the CS. It also encompasses the PHD method to estimate the CMS and provides several helper functions to estimate additional model parameters in each case. The \textbf{MAVE} R package, developed by \cite{Weiq2019}, implements the MAVE and OPG methods to estimate the CS and CMS. The R package \textbf{orthoDr}, developed by \cite{orthodr2019}, provides estimates of the CS using the Eff and pSAVE methods. Finally, the \textbf{nsdr} R package, developed by \cite{nsdr2021}, implements the KPCA, GSIR, and GSAVE methods to estimate the CS. 

Our R package, \textbf{itdr}, performs the latest developments in the estimation of the CS and CMS using integral transformation methods. The \textbf{itdr} provides functions to estimate both the CS and CMS using various approaches. These include the FM approach proposed by \cite{zhu2006} under the normality assumption of the predictors, the CM approach introduced by \cite{zeng2010}, and the IHT method by \cite{CL02}.
Furthermore, \cite{zeng2010} established a general framework for any integral transformation method that can be used to estimate the CS and CMS when the predictor vector follows either a multivariate normal distribution, an elliptically contoured distribution, or without imposing any distributional assumption on it.  In this framework, a kernel smoother is applied to approximate the unknown distribution function of the predictor variables.
Our \textbf{itdr} package also includes some helper functions to estimate the dimension of SDR subspaces and the tuning parameters of the FM and CM algorithms. The \textbf{itdr} package facilitates the estimation of the CS using the inverse regression approach via Fourier transformation (invFM) proposed by \cite{Weng18}, the minimum discrepancy approach\citep{weng2022minimum}, and the sparse Fourier transform inverse regression method \citep{weng2022fourier}. These Fourier transformation methods allow for using either univariate or multivariate responses and provide an estimate for the CS. 
The R package \textbf{itdr} has been uploaded to CRAN and is available at \url{https://CRAN.R-project.org/package=itdr}.

The remainder of this paper is organized as follows; we start with a brief overview of six integral transformation methods employed for estimating the SDR subspaces. Once the methods are introduced, the subsequent sections are devoted to demonstrating the utilization of functions within  the \textbf{itdr} package. Specifically, we illustrate how to estimate the tuning parameters, select the dimension of SDR subspaces, and obtain accurate estimations of the SDR subspaces using various datasets available within  the \textbf{itdr} package.


\section{Integral Transformation Methods (ITMs) and Candidate Matrices}

In this section, we provide a summary of the theoretical framework of integral transformation methods proposed by \cite{zhu2006} and \cite{zeng2010}. These methods are utilized to drive estimators for the SDR subspaces. All of these methods employ the spectral-decomposition-based procedure. The first step of each method involves constructing a nonnegative definite symmetric matrix, denoted as $\mathbf{M} \in \mathbb{R}^{p \times p}$, referred to as a candidate matrix. The second step consists of performing the spectral-decomposition of a sample version of $\mathbf{M}$, denoted as  $\widehat{\mathbf{M}}$. Then, an orthogonal basis is formed by the first $d$ eigenvectors corresponding to the leading $d$ eigenvalues of $\widehat{\mathbf{M}}$, which serves as an estimation of the target SDR subspace. The candidate matrices based on the integral transformation method, whose column spaces are identical to the CMS and CS, are respectively denoted as $\mathbf{M}_{ITM}$, and $\mathbf{M}_{ITC}$.

First, we consider the derivation of the candidate matrix $\mathbf{M}_{ITM}$ for estimating the CMS in regression. Suppose $m(\mathbf{x})=E(Y\vert \mathbf{X}=\mathbf{x})$; then from the model given in Equation \eqref{eq:1}, we can express $m(\mathbf{x})$ as $m(\mathbf{x})=g(\boldsymbol{\Gamma}^T\mathbf{x})$. By applying the chain rule of differentiation, we establish the following relationship between the gradient operator of $m(\mathbf{x})$ and the link function $g(\mathbf{u})$, where $\mathbf{u}=\boldsymbol{\Gamma}^T\mathbf{x}$,
\begin{equation}
	\frac{\partial}{\partial \mathbf{x}}m(\mathbf{x})=\boldsymbol{\Gamma}\frac{\partial}{\partial \mathbf{u}}g(\mathbf{u}).
\end{equation}
Since $\boldsymbol{\Gamma}\frac{\partial}{\partial \mathbf{u}}g(\mathbf{u}) \in \mathcal{S}_{E[Y \vert \mathbf{X}]}$, it follows that  $\frac{\partial}{\partial \mathbf{x}}m(\mathbf{x}) \in \mathcal{S}_{E[Y\vert \mathbf{X}]} $.  This further implies that  $E\left[\frac{\partial}{\partial \mathbf{X}}m(\mathbf{X})\right] \in \mathcal{S}_{E[Y\vert \mathbf{X}]}$  providing the average derivative estimate (ADE). However, the ADE method can only generate one direction, limiting its ability to estimate the CMS of dimensions higher than one. Additionally, the ADE method fails when $E\left[\frac{\partial}{\partial \mathbf{X}}m(\mathbf{X})\right]=0$. More detailed discussions can be found in \cite{zeng2010}. These disadvantages can be overcome by introducing an appropriate family of weight functions $\{W(\mathbf{x},\mathbf{u}): \mathbf{u} \in \mathbb{R}^p\}$, such that $E\left[W(\mathbf{x},\mathbf{u})\frac{\partial}{\partial \mathbf{X}}m(\mathbf{X})\right]\neq 0$,  where $\mathbf{u} \in \mathbb{R}^{p}$ represents the family index. 
Note that a single weight function is sufficient to overcome the first drawback; however, different weight functions produce different vectors in the CMS, enabling the  estimation of   the entire CMS. The weighted ADE is then defined as
\begin{equation}\label{eq:5}
	\boldsymbol{\psi}(\mathbf{u})=E\left[\frac{\partial}{\partial \mathbf{X}}m(\mathbf{X})W(\mathbf{X},\mathbf{u})\right]=\int \frac{\partial}{\partial \mathbf{x}}m(\mathbf{x})W(\mathbf{x},\mathbf{u}) f(\mathbf{x})d\mathbf{x}.
\end{equation}
Indeed, $\boldsymbol{\psi}(\mathbf{u})$ represents the integral transformation of the density weighted gradient of $m(\mathbf{x})$, and $W(\mathbf{x},\mathbf{u})$ is the nondegenerate kernel function associated with this transformation. The most commonly chosen weight function are $W_1(\mathbf{x},\mathbf{u})=\exp\{i\mathbf{u}^T\mathbf{x}\}$ and $W_2(\mathbf{x},\mathbf{u})=H(\mathbf{u}-\mathbf{x})$,  where $H(\cdot)$ is an absolutely integrable function. The integral transformation using the weigh function $W_1(\mathbf{x},\mathbf{u})$ is referred to as the Fourier transformation method (FM), while   the transformation using the weight function $W_2(\mathbf{x},\mathbf{u})$ is known as the convolution transformation method (CM). By applying the integration by parts to Equation \eqref{eq:5}, we obtain
\begin{equation}\label{eq:6}
	\boldsymbol{\psi}(\mathbf{u})=-E\left[Y\phi(\mathbf{X},\mathbf{u})\right],
\end{equation}
where $\phi(\mathbf{x},\mathbf{u})=\frac{\partial}{\partial \mathbf{x}}W(\mathbf{x},\mathbf{u})+W(\mathbf{x},\mathbf{u})\mathbf{g}(\mathbf{x})$ and $\mathbf{g}(\mathbf{x})=\frac{\partial}{\partial \mathbf{x}}\log f(\mathbf{x})$. Note that the expression  $\boldsymbol{\psi}(\mathbf{u})$ in Equation \eqref{eq:5} depends on the mean function $m(\mathbf{x})$,  while  $\boldsymbol{\psi}(\mathbf{u})$ in Equation \eqref{eq:6} does not rely on $m(\mathbf{x})$.  This implies that we can calculate  $\boldsymbol{\psi}(\mathbf{u})$  without the need to fit or  estimate the link function $g(\cdot)$ or its derivatives. Therefore,   $\boldsymbol{\psi}(\mathbf{u})$ in Equation \eqref{eq:6} is used to define a candidate matrix for $\mathcal{S}_{E[{Y\vert \mathbf{X}}]}$. This candidate matrix,  whose column space is equal to the CMS, is denoted by $\mathbf{M}_{ITM}$ and defined as follows
\begin{equation}\label{eq:7}
	\mathbf{M}_{ITM}=\int \boldsymbol{\psi}(\mathbf{u})\overline{\boldsymbol{\psi}}^T(\mathbf{u}) K(\mathbf{u})d\mathbf{u}=E[\mathcal{\mathbf{U}}_{ITM}(\mathbf{Z}_1,\mathbf{Z}_2)],
\end{equation}
where $\overline{\boldsymbol{\psi}}(\cdot)$ denotes the conjugation of $\boldsymbol{\psi}(\cdot)$, $\mathbf{z}=(y,\mathbf{x})$, $\mathbf{Z}_1$ and $\mathbf{Z}_2$ are two independent realizations of $\mathbf{Z}$, and
\begin{equation}\label{eq:8}
	\mathcal{\mathbf{U}}_{ITM}(\mathbf{z}_1,\mathbf{z}_2)=y_1y_2\int \phi(\mathbf{x}_1,\mathbf{u})\phi^T(\mathbf{x}_2,\mathbf{u}) K(\mathbf{u}) d\mathbf{u}.
\end{equation}
Note that for the FM method, $K(\mathbf{u})$ is set as $K(\mathbf{u})=(2\pi \sigma_u^2)^{-p/2}\exp\{-\vert \vert \mathbf{u}\vert \vert/2\sigma_u^2\}$, and for the CM it is set to $K(\mathbf{u})=1$. Let $\mathcal{S}(\mathbf{M}_{ITM})$ denote the subspace spanned by the columns of $\mathbf{M}_{ITM}$. Then, Lemma \ref{lem:2} establishes that the column space of $\mathbf{M}_{ITM}$ is the same as the CMS.  
\begin{Lemma}\label{lem:2}
	 The result stated in Equation \eqref{eq:6} holds under the following conditions:	 
	 \begin{itemize}	 	
	 \item [a.]  The function $f(\mathbf{x})\frac{\partial}{\partial \mathbf{x}}m(\mathbf{x})$ exists and is absolutely integrable.
	 \item [b. ]   The expression $W(\mathbf{x},\mathbf{u})m(\mathbf{x})f(\mathbf{x})$ goes to zero as $\vert \vert \mathbf{x}\vert \vert \to \infty$. 
	 \item [c.]  The function $\log f(\mathbf{x})$ is differentiable. 	 
	\end{itemize}
Moreover, if $W(\cdot,\cdot)$ is a nondegenerate kernel for $f(\mathbf{x})\frac{\partial}{\partial \mathbf{x}}$ and $\boldsymbol{\psi}(\mathbf{u})$ is square integrate, then $\mathbf{M}_{ITM}$ is a nonnegative definite matrix and $\mathcal{S}(\mathbf{M}_{ITM})=\mathcal{S}_{E[Y\vert \mathbf{X}]}$.
\end{Lemma}
\textcolor{black}{The proof of Lemma 1 is similar to that of Proposition 2 in \cite{zhu2006} and Lemma 3 in \cite{zeng2010}, and hence is omitted. }
Furthermore, a candidate matrix for the CS can be obtained by following the procedures outlined below. To derive it, it is important to note that the central mean subspace is always a subspace of the central subspace, represented as  $\mathcal{S}_{E[Y\vert \mathbf{X}]} \subseteq \mathcal{S}_{Y \vert \mathbf{X}}$ \cite[][]{cook98}. Let $T(Y)$ denote an arbitrary transformation of the response variable $Y$,  resulting in a new response. 
Furthermore, let $\mathcal{S}_{E[T(Y)\vert \mathbf{X}]}$ denotes the CMS of $T(Y)$ on $\mathbf{X}$.  It is  then evident that $\mathcal{S}_{E[T(Y)\vert \mathbf{X}]} \subseteq \mathcal{S}_{Y\vert \mathbf{X}}$,  but it 
may not be identical to $\mathcal{S}_{E[Y\vert \mathbf{X}]}$ \cite[][]{zhu2006}. 
Despite the fact that   two CMS $\mathcal{S}_{E[T_1(Y)\vert \mathbf{X}]}$ and $\mathcal{S}_{E[T_2(Y)\vert \mathbf{X}]}$ 
differ for two distinct transformations $T_1(\cdot)$ and $T_2(\cdot)$,  they might  cover different parts of the entire CS.  Thus, by considering a collection of  CMS obtained from different transformations, it becomes possible to cover the entire CS \cite[][]{zhu2006}. In other words, we have 
\begin{equation*}
	\mathcal{S}_{Y\vert \mathbf{X}}=\sum_{all~~possible~~ T}\mathcal{S}_{E[T(Y)\vert \mathbf{X}]}.
\end{equation*}
Let $T_v(Y)$ represent a simple family of transformations, denoted as  $\{T_v(\cdot): T_v(y)=\tau(y,v), \text{ for }y,v\in \mathbb{R}\}$, where $\tau(\cdot,\cdot)$ is a known function. For a given $v \in \mathbb{R}$, the conditional mean response of $T_v(Y)$  given $\mathbf{x}$ is defined as $m(\mathbf{x},v)=E[\tau(Y,v)\vert \mathbf{X}=\mathbf{x}]$. Now, let us define the integral transformation of  the partial derivative of $m(\mathbf{x},v)$  with respect to $\mathbf{x}$,  
weighted by $W(\mathbf{x},\mathbf{u})$ as $\boldsymbol{\psi}(\mathbf{u},v)$.  In other words 
\begin{equation}\label{eq:CS}
	\boldsymbol{\psi}(\mathbf{u},v)=E\left[\frac{\partial m(\mathbf{X},v)}{\partial \mathbf{x}}W(\mathbf{x},\mathbf{u})\right]=-E\left[\tau(Y,v)\phi(\mathbf{X},\mathbf{u})\right].
\end{equation}
Furthermore, a candidate matrix for $\mathcal{S}_{Y\vert \mathbf{X}}$, denoted as $\mathbf{M}_{ITC}$, is defined as
\begin{equation}\label{eq:7.1}
\mathbf{M}_{ITC}=\int \boldsymbol{\psi}(\mathbf{u},v)\boldsymbol{\psi}^T(\mathbf{u},v) K(\mathbf{u})d\mathbf{u}=E[\mathbf{U}_{ITC}(\mathbf{Z}_1,\mathbf{Z}_2)],
\end{equation}
 where $K(\mathbf{u})=(2\pi\sigma_u^2)^{-p/2}\exp\{-||\mathbf{u}||/(2\sigma_u^2)\}$ for the FM procedure and $K(\mathbf{u})=1$ for the CM procedure. In Equation \eqref{eq:7.1}, $\mathbf{z}=(y,\mathbf{x})$, $\mathbf{Z}_1$ and $\mathbf{Z}_2$ represent independent realizations of $\mathbf{Z}$, and
\begin{equation}\label{eq:8.1}
\mathbf{U}_{ITC}(\mathbf{z}_1,\mathbf{z}_2)=\int\tau(y_1,v)\tau(y_2,v)dv \int \phi(\mathbf{x}_1,\mathbf{u})\phi^T(\mathbf{x}_2,\mathbf{u}) K(\mathbf{u})d\mathbf{u},
\end{equation}
where $\tau(y,v)=(2\pi\sigma_v^2)^{-1/2}\exp\{-(y-v)^2/(2\sigma_v^2)\}$. Lemma \ref{lemma2} demonstrates that under certain mild  conditions $\mathcal{S}_{Y\vert \mathbf{X}}=\sum_{v\in \mathbb{R}} \mathcal{S}_{E[T_v(Y)\vert \mathbf{X}]}$, and the column space of $\mathbf{M}_{ITC}$ in Equation \eqref{eq:7.1} is the same as the central mean subspace $\mathcal{S}_{Y\vert\mathbf{X}}$. 
\begin{Lemma}\label{lemma2}
	The results presented in Equation \eqref{eq:CS} are valid under the folloing conditions: 
	\begin{itemize}
	\item [a. ] $f(\mathbf{x})\frac{\partial}{\partial \mathbf{x}}m(\mathbf{x},v)$ exists and is absolutely integrable.
	\item[b. ] $W(\mathbf{x},\mathbf{u})m(\mathbf{x},v)f(\mathbf{x})$ goes to zeros as $||\mathbf{x}|| \to \infty$ .
	\item[c. ]$\log f(\mathbf{x})$ is differentiable.
\end{itemize}
Furthermore, if $\tau(y,v)W(\mathbf{x},\mathbf{u})$ serves as  nondegenerate kernal for $f(\mathbf{x})\frac{\partial}{\partial \mathbf{x}}$ and $\boldsymbol{\psi}(\mathbf{u},v)$ is square integrable, then $\mathbf{M}_{ITC}$ is a nonnegative definite matrix and $\mathcal{S}(\mathbf{M}_{ITC})=\mathcal{S}_{Y\vert \mathbf{X}}$.
\end{Lemma}
 The proof of Lemma \ref{lemma2} follows a  similar approach to that of Proposition 6 in \cite{zhu2006} and Lemma 4 in \cite{zeng2010}, and is therefore omitted here.

\subsection{Estimation of Candidate Matrices}
 In this section,  we discuss the estimates of the candidate matrices $\mathbf{M}_{ITM}$ and $\mathbf{M}_{ITC}$.    We begin by expressing the sample versions of the candidate matrices   used to estimate the CMS and the CS under the Fourier transformation method. We denote these candidate matrices as $\mathbf{M}_{FMM}$ and $\mathbf{M}_{FMC}$, respectively. First, we set $W(\mathbf{x},\mathbf{u})=\exp\{i\mathbf{u}^T\mathbf{x}\}$ in Equation \eqref{eq:5}. Then, the sample version of the candidate matrix under the Fourier transformation method,  which targets the CMS, can be expressed as   
 \begin{equation}\label{eq:12}
 	\widehat{\mathbf{M}}_{FMM}^{\star}=\frac{1}{N^2}\sum_{i=1}^N\sum_{j=1}^N \widehat{\mathbf{U}}_{FMM}(\mathbf{Z}_i,\mathbf{Z}_j),
 \end{equation}
 where
 \begin{equation}\label{eq:13}
 	\widehat{\mathbf{U}}_{FMM}(\mathbf{Z}_1,\mathbf{Z}_2)=y_1y_2\exp\left\{\frac{-\sigma_u^2}{2}||\mathbf{u}_{12}||^2\right\}\left[\sigma_u^2\mathbf{I}_p+[{\mathbf{g}}(\mathbf{x}_1)-\sigma_u^2\mathbf{u}_{12}][{\mathbf{g}}(\mathbf{x}_2)+\sigma_u^2\mathbf{u}_{12}]^T\right].
 \end{equation}
 Here ${\mathbf{g}}(\mathbf{x})=\frac{\frac{\partial}{\partial \mathbf{x}}{f}(\mathbf{x})}{{f}(\mathbf{x})}$, and $\mathbf{u}_{12}=\mathbf{x}_1-\mathbf{x}_2$. Note that the density function ${f}(\mathbf{x})$ appears in the denominator of ${\mathbf{g}}(\mathbf{x})$. Therefore, to mitigate the negative effect of small values of ${f}(\mathbf{x})$ on the estimation, the estimator $\widehat{\mathbf{M}}_{FMM}^{\star}$ given in Equation \eqref{eq:12} is modified as follows
 \begin{equation}\label{eq:14}
 	\widehat{\mathbf{M}}_{FMM}=\frac{1}{N^2}\sum_{i=1}^N\sum_{j=1}^N \widehat{\mathbf{U}}_{FMM}(\mathbf{Z}_i,\mathbf{Z}_j)\widehat{I}_i\widehat{I}_j,
 \end{equation}
 where $\widehat{\mathbf{U}}_{FMM}$ is the same as  in Equation \eqref{eq:13}. Additionally, for a predetermined threshold value $b$, we defined $\widehat{I}_{(\cdot)}=1$ if ${f}(\mathbf{x}_{(\cdot)})>b$ and $\widehat{I}_{(\cdot)}=0$ otherwise. Similarly, by setting $\tau(y,v)=(2\pi\sigma_v^2)^{-1/2}\exp\{-(y-v)^2/(2\sigma_v^2)\}$, the sample version of the candidate matrix $\mathbf{M}_{FMC}$  targeting the CS is given as
 \begin{equation}\label{eq:15}
 	\widehat{\mathbf{M}}_{FMC}=\frac{1}{N^2} \sum_{i=1}^{N}\sum_{j=1}^{N}\widehat{\mathbf            {U}}_{FMC}(\mathbf{Z}_i,\mathbf{Z}_j)\widehat{I}_i\widehat{I}_j,
 \end{equation}
where
\begin{equation}\label{eq:16}
\widehat{\mathbf{U}}_{FMC}(\mathbf{Z}_1,\mathbf{Z}_2)=\exp\left\{\frac{-\sigma_u^2}{2}||\mathbf{u}_{12}||^2-\frac{\sigma_v^2}{2}v^2_{12}\right\}\left[\sigma_u^2\mathbf{I}_p+[\mathbf{g}(\mathbf{x}_1)-\sigma_u^2\mathbf{u}_{12}][\mathbf{g}(\mathbf{x}_2)+\sigma_u^2\mathbf{u}_{12}]^T\right],
\end{equation}
where $\mathbf{u}_{12}=\mathbf{x}_1-\mathbf{x}_2$, and $v_{12}=y_1-y_2$. 

Now, we express the sample versions of the candidate matrices for estimating the CMS and CS under the convolution transformation method (CM). These candidate matrices are denoted as $\mathbf{M}_{CMM}$ and $\mathbf{M}_{CMC}$, respectively.  Their column spaces are identical to the CMS and the CS. For both the CMS and CS,  we set the same weight function $W(\mathbf{x},\mathbf{u})=H(\mathbf{x}-\mathbf{u})$ in Equation \eqref{eq:5}, where $H(\mathbf{x}-\mathbf{u})=(2\pi\sigma_u^2)^{-p/2}\exp\{(\mathbf{x}-\mathbf{u})^T(\mathbf{x}-\mathbf{u})/(2\sigma_u^2)\}$.  By applying simple algebra manipulations, we obtain
\begin{equation}\label{eq:17}
	\widehat{\mathbf{M}}_{CMM}=\frac{1}{N^2}\sum_{i=1}^{N}\sum_{j=1}^{N} \widehat{\mathbf{U}}_{CMM}(\mathbf{Z}_i,\mathbf{Z}_j)\widehat{I}_i\widehat{I}_j,
\end{equation}
where
\begin{equation}\label{eq:18}
		\widehat{\mathbf{U}}_{CMM}(\mathbf{Z}_1,\mathbf{Z}_2)=y_1y_2\exp\left\{\frac{-||\mathbf{u}_{12}||^2}{4\sigma_u^2}\right\}\left[\frac{1}{2\sigma_u^2}\mathbf{I}_p+[\mathbf{g}(\mathbf{x}_1)-\frac{1}{2\sigma_u^2}\mathbf{u}_{12}][\mathbf{g}(\mathbf{x}_2)+\frac{1}{2\sigma_u^2}\mathbf{u}_{12}]^T\right].
\end{equation}
 
Similarly, we choose $\tau(y,v)=\exp \{-(y-v)^2/(2\sigma_v^2)\}$ to derive an expression for the candidate matrix $\mathbf{M}_{CMC}$ that targets the CS. 
The sample version of $\mathbf{M}_{CMC}$ is given by
\begin{equation}\label{eq:19}
	\widehat{\mathbf{M}}_{CMC}=\frac{1}{N^2}\sum_{i=1}^{N}\sum_{j=1}^{N} \widehat{\mathbf{U}}_{CMC}(\mathbf{Z}_i,\mathbf{Z}_j)\widehat{I}_i\widehat{I}_j,
\end{equation}
where
\begin{equation}\label{eq:20}
\widehat{\mathbf{U}}_{CMC}(\mathbf{Z}_1,\mathbf{Z}_2)=\exp\left\{-\frac{v_{12}^2}{4\sigma_v^2}- \frac{||\mathbf{u}_{12}||^2}{4\sigma_u^2}\right\}\left[\frac{1}{2\sigma_u^2}\mathbf{I}_p+[\mathbf{g}(\mathbf{x}_1)-\frac{1}{2\sigma_u^2}\mathbf{u}_{12}][\mathbf{g}(\mathbf{x}_2)+\frac{1}{2\sigma_u^2}\mathbf{u}_{12}]^T\right].
\end{equation}

\subsection{Estimating Density Functions}
 The density function $f(\mathbf{x})$ which appears in the term $\mathbf{g}(\mathbf{x})=\frac{\frac{\partial }{\partial \mathbf{x}}f(\mathbf{x})}{f(\mathbf{x})}$ is unknown. In this section, we discuss the methods that are used to approximate the unknown density function $f(\mathbf{x})$ at a given point $\mathbf{x}$.  One approach is to  assume $f(\mathbf{x})=f_0(\mathbf{x}; \boldsymbol\theta)$, where $f_0$ is a known function and $\boldsymbol\theta$ is a vector of unknown parameters. By making this assumption,  we can estimate $f(\mathbf{x})$ and $g(\mathbf{x})$ parametrically. A well-known parametric density is the Gaussian density function,  which can be considered for $f(\mathbf{x})$. Under the normality assumption, it can be shown that $g(\mathbf{x})=-\mathbf{x}$. 
 
 Alternatively,   we can employ a nonparametric method to   estimate   $f(\mathbf{x})$ without imposing distributional assumptions on $\mathbf{X}$. In this setting, \cite{zeng2010} utilized kernel density estimation approach to estimate $f(\mathbf{x})$. They proposed the following estimator for $\mathbf{g}(\mathbf{x})$ at a specific point $\mathbf{x}_0$
 \begin{equation}
 	\widehat{\mathbf{g}}(\mathbf{x}_0)=\frac{\frac{\partial}{\partial \mathbf{x}_0}\widehat{f}(\mathbf{x}_0)}{\widehat{f}(\mathbf{x}_0)}=\frac{(nh^{p+1})^{-1}\sum_{\ell=1}^{n}K^{'}((\mathbf{x}_0-\mathbf{x}_{\ell})/h)}{(nh^{p})^{-1}\sum_{\ell=1}^{n} K((\mathbf{x}_0-\mathbf{x}_{\ell})/h)},
 \end{equation}
 where $\widehat{f}(\mathbf{x}_0)=(nh^{p})^{-1}\sum_{\ell=1}^{n} K((\mathbf{x}_0-\mathbf{x}_{\ell})/h)$, $K(\cdot)$ is a kernel function with bandwidth $h$, and $K^{'}(\cdot)$ represents the derivative of $K(\cdot)$. 
 
 Moreover, another option is to assume an elliptically contoured distribution for $f(\mathbf{x})$. It has been shown that, under an elliptically contoured distribution, an estimate for $g(\mathbf{x})$ can be  obtained as follows \cite[for more details, see][]{zeng2010}
 \begin{equation}
 	\widetilde{\mathbf{g}}(\mathbf{x}_i)=\frac{\mathbf{x}_i}{r_i}\frac{\widetilde{f}'(r_i)}{\widetilde{f}(r_i)}-\frac{p-1}{r_i^2}\mathbf{x}_i,
 \end{equation}
 where $r_i=||\mathbf{x}_i||$, and $\{r_i\}_{i=1}^n$ represents an independent and identically distributed (iid) sample from $f(r)$. Additionally, $\widetilde{f}(\cdot)$ denotes the kernel density estimation of $f(\cdot)$ with bandwidth $h$, and $\widetilde{f}^{'}(\cdot)$ is the derivative of $\widetilde{f}(\cdot)$. We use  $\widehat{\mathbf{M}}_{FMMn}$ to denote the estimate of the candidate matrix  $\mathbf{M}_{FMM}$ under the Gaussian assumption, $\widehat{\mathbf{M}}_{FMMk}$ for the estimate of $\mathbf{M}_{FMM}$ under the kernel density smoother, and $\widehat{\mathbf{M}}_{FMMe}$ as the corresponding estimator under the
 elliptically contoured distribution. Table \ref{tab:1} provides a summary of the estimators of the candidate matrices for the CMS and the CS under different assumptions. 

\begin{table}[h!]
	\centering
	\footnotesize
	\caption{Estimators of the candidate matrices under different density assumptions based on two transformation methods (FM and CM)}\label{tab:1}
	\begin{tabular}{|c|c|c|c|c|}
	\hline
	Method &SDR subspace&Assumption&Estimator &Replace $g(\mathbf{x})$ by\\  \hline
	&&&&\\
 	\multirow{3}{*}{Fourier Transformation}&\multirow{3}{*}{CMS (CS)}&Normal&$\widehat{\mathbf{M}}_{FMMn} (\widehat{\mathbf{M}}_{FMCn})$&	 $-\mathbf{x}$\\
	&&Kernel&$\widehat{\mathbf{M}}_{FMMk}(\widehat{\mathbf{M}}_{FMCk})$&	 $\widehat{g}(\mathbf{x})$\\
	&&Elliptical&$\widehat{\mathbf{M}}_{FMMe}(\widehat{\mathbf{M}}_{FMCe})$&	 $\widetilde{g}(\mathbf{x})$\\ &&&&\\
	\hline
	&&&&\\
	\multirow{3}{*}{Convolution Transformation}&\multirow{3}{*}{CMS (CS)}&Normal&$\widehat{\mathbf{M}}_{CMMn} (\widehat{\mathbf{M}}_{CMCn})$&	 $-\mathbf{x}$\\
	&&Kernel&$\widehat{\mathbf{M}}_{CMMk}(\widehat{\mathbf{M}}_{CMCk})$&	 $\widehat{g}(\mathbf{x})$\\
	&&Elliptical&$\widehat{\mathbf{M}}_{CMMe}(\widehat{\mathbf{M}}_{CMCe})$& $\widetilde{g}(\mathbf{x})$\\ 
	&&&&\\
	\hline
	\end{tabular}
\end{table}
\cite{zhu2006} proposed an algorithm for estimating the candidate matrices of the CMS  and the CS based on the Fourier transformation method, assuming normality  for the predictors. These estimators are denoted  as $\widehat{\mathbf{M}}_{FMMn}$ and $\widehat{\mathbf{M}}_{FMCn}$, respectively. When employing the kernel density smoother and assuming an elliptical density, a similar algorithm can be applied to estimate these candidate matrices by replacing $g(\mathbf{x})$ with $\widehat{g}(\mathbf{x})$ and $\widetilde{g}(\mathbf{x})$, respectively. Furthermore, the estimators based on the convolution transformation method can also be obtained using the same algorithm. The details of this algorithm are summarized in Algorithm 1.

Let $\{Y_i,\mathbf{X}_i\}_{i=1}^n$ denote an iid sample of size $n$ from $(Y,\mathbf{X})$. Furthermore, assume $Cov(\mathbf{X})=\boldsymbol\Sigma$, and suppose the known dimension of the SDR subspace $(d)$, as well as   the known tuning parameters $\sigma_u^2$ and $\sigma_v^2$. Algorithm \ref{algo:1} outlines the steps for estimating the CMS and the CS. 

\begin{algorithm}[H]
	\label{algo:1}
	\SetAlgoLined
	\begin{enumerate}
        \item [1. ] Standardize the response variable and predictor variables as follows: $\widetilde{\mathbf{x}}_i=\widehat{\boldsymbol\Sigma}^{-1/2}(\mathbf{x}_i-\overline{\mathbf{x}})$ and $\widetilde{y}_i=s_y^{-1}(y_i-\overline{y})$, where $\overline{\mathbf{x}}$ and $\widehat{\boldsymbol\Sigma}$ are the sample mean and the sample covariance matrix of the $\mathbf{x}_i$'s, and $\overline{y}$ and $s_y$ are the sample mean and standard deviation of the response variable.
 	\item [2. ] Obtain $\widehat{\mathbf{M}}_{FMMn}$ ( $\widehat{\mathbf{M}}_{FMCn}$) using the standardized data $\{(\widetilde{y}_i,\widetilde{\mathbf{x}}_i)\}_{i=1}^n$.
 	\item [3. ]Calculate the eigen decomposition of $\widehat{\mathbf{M}}_{FMMn}$ ( $\widehat{\mathbf{M}}_{FMCn}$), and let $(\widehat{\mathbf{e}}_1,\widehat{\lambda}_1), \cdots, (\widehat{\mathbf{e}}_p,\widehat{\lambda}_p)$ be the corresponding eigenvector-eigenvalue pairs with $\widehat{\lambda}_1 \geq\cdots\geq \widehat{\lambda}_p$.
 	\item [4. ] Estimate the CMS (CS) by calculating $\widehat{\mathcal{S}}_{E[y\vert \mathbf{x}]}$ ($\widehat{\mathcal{S}}_{y\vert \mathbf{x}}$)=Span$\left\{\widehat{\boldsymbol{\Sigma}}^{-1/2}\widehat{\mathbf{e}}_1,\cdots,\widehat{\boldsymbol{\Sigma}}^{-1/2}\widehat{\mathbf{e}}_d \right\}$.
	\end{enumerate}
	\caption{Estimation of the CSM and the CS}
\end{algorithm}


\subsection{Estimating the Dimension of the SDR Subspace}
The dimension of the SDR subspace can be estimated using a bootstrap sampling procedure.  A metric is employed to measure the distance between two subspaces. Consider $\mathcal{S}_1$ and $\mathcal{S}_2$ as two subspaces spanned by the columns of two matrices $\mathbf{A}$ and $\mathbf{B}$, respectively, with full column rank.  The distance between $\mathcal{S}_1$ and $\mathcal{S}_2$ is defined based on the trace correlation $(\gamma)$ as   $\gamma=\sqrt{(1/d)\text{tr}(\mathbf{P}_{\mathbf{A}}\mathbf{P}_{\mathbf{B}})}$, where $\mathbf{P}_{\mathbf{A}}=\mathbf{A}(\mathbf{A}^T\mathbf{A})^{\dagger}\mathbf{A}^T$, $\mathbf{P}_{\mathbf{B}}=\mathbf{B}(\mathbf{B}^T\mathbf{B})^{\dagger}\mathbf{B}^T$, $\dagger$ denotes the generalized inverse of a matrix, and $\mathbf{P}_{(\cdot)}$ represents the projection matrix. The range of $\gamma$ is  $0 \leq \gamma \leq 1$, with $\gamma=1$ when $\mathcal{S}_1$ and $\mathcal{S}_2$ are identical, and $\gamma=0$ indicating $\mathcal{S}_1$ and $\mathcal{S}_2$ are perpendicular. Therefore, the metric $D=1-\gamma$ can be employed to measure the distance between $\mathcal{S}_1$ and $\mathcal{S}_2$. 
To estimate the dimension $d$, let $\{Y_i,\mathbf{X}_i\}_{i=1}^n$ be a random sample of size $n$ from $(Y,\mathbf{X})$. The following algorithm can be used for estimating the dimension. Algorithm \ref{algo:2} outlines the steps for selecting the dimension of the SDR subspace.

\begin{algorithm}[H]
	\label{algo:2}
	\SetAlgoLined
	\begin{enumerate}
        \item [1.] Randomly generate $B$ bootstrap samples of size $n$ from the original sample $\{y_i,\mathbf{x}_i\}_{1\leq i \leq B}$ with replacement. Denote the $b$th bootstrap sample as $\{y_i^{(b)},\mathbf{x}_i^{(b)}\}$, where $b=1,\dots,B$. 
	
	\item [2.] For each bootstrap sample, i.e.,$\{y_i^{(b)},\mathbf{x}_i^{(b)}\}$, obtain the estimator for the SDR subspace using Algorithm \ref{algo:1} and denote it as $\widehat{\mathcal{S}}_{d}^{(b)}$. 
	
	\item [3. ] Calculate the distance between $\widehat{\mathcal{S}}_{d}^{(b)}$ and $\widehat{\mathcal{S}}_d$, where $\widehat{\mathcal{S}}_d$ represents the estimated SDR subspace from the original sample. Denote this distance as $D_q^{(b)}$.
	
	\item [4.] Repeat steps 2 and 3 for each bootstrap sample. Then, calculate the mean distance between $\widehat{\mathcal{S}}_{d}^{(b)}$ and $\widehat{\mathcal{S}}_d$ over all bootstrap samples $b=1,\dots,B$. That is,
	\begin{equation}
	\overline{D}(d)=\frac{1}{B}\sum_{b=1}^B D_{d}^{(b)}.
	\end{equation}   
	\end{enumerate}
 Repeat  steps 1 to 4, for $d=1,\dots,p$, and obtain the sequence of average distances $\{\overline{D}(d)\}_{d=1}^p$.
	\caption{Subspace dimension selection}
\end{algorithm}
 
 This sequence serves as a  measure of variability for $\widehat{\mathcal{S}}_d$. According to \cite{zhu2006}, an estimate of $d$ is obtained as follows.
Plot $\overline{D}(d)$ verses $d$, creating a dimension variability plot. 
In the next step, analyze the overall trend in the plot while disregarding the local fluctuations that do not align with the overall trend. Suppose the estimator for $d$ is $d_0$.  In this case,  $\overline{D}(d)$ decreases for $1 \leq d \leq d_0$, and increases for $d_0 \leq d \leq d^{\star}$, where $d^{\star}$ is  the value that maximizes $\overline{D}(d)$. Subsequently, $\overline{D}(d)$ decreases to zero for $d^{\star}\leq d\leq p$. The value of $d_0$ is referred to as the valley point and $d^{\star}$ is referred to as the peak of the trend.  
  
\subsection{Estimating the Tuning Parameters}
There are three model parameters needed to be tuned before starting the SDR estimation procedure described in Algorithms \ref{algo:1}.   However, not all three parameters need to be tuned for each method. Specifically, only the tuning parameter $\sigma^2_{u}$ is required when estimating the CMS, while both tuning parameters $\sigma^2_{u}$ and $\sigma^2_{v}$ are required when estimating the CS. Additionally, the tuning parameter $h$ is only necessary when using the kernel density smoother to estimate the density function of the predictor variables. \cite{zhu2006} proposed the following procedure to estimate the tuning parameters. First, find the optimal subspace dimension $d$, denoted as $d_0$, by applying the bootstrap procedure described in Algorithm \ref{algo:2} with $\sigma^2_{u}=0.1$, and $\sigma^2_v=1.0$. 
Second, determine the value of  $\sigma^2_{u}$, denoted as $\sigma^2_{u_1}$,  by utilizing Algorithm \ref{algo:3}   with $d=d_0$ and $\sigma^2_{v}=1.0$. 
Third, similarly choose the valued of  $\sigma_v^2$, denoted as   $\sigma^2_{v_1}$,  by setting $d=d_0$ and $\sigma^2_u=\sigma^2_{u1}$. 
Finally, estimate the valued of  $h$, denoted as $h_0$,  by setting $d=d_0$, $\sigma_u^2=\sigma^2_{u_1}$, and $\sigma_v^2=\sigma^2_{v_1}$. The bootstrap estimation procedure for estimating all three tuning parameters is the same in each case and is outlined in Algorithm \ref{algo:3} for estimating $\sigma^2_{u}$. A similar procedure can be employed to select optimal values for $\sigma^2_{v}$ and $h$.

Suppose $\sigma^2_{u,1},\dots, \sigma^2_{u,k}$ represents a set of candidate values for $\sigma_u^2$ that are equally spaced within a given interval. Then, repeat the following algorithm for each candidate value $\sigma^2_{u,\ell}$, $\ell=1,\dots,k$, to obtain a sequence of the mean distances referred to as the variability measure, denoted as $\{\overline{D}(\sigma_{u,\ell}^2)\}_{\ell=1}^k$.

\begin{algorithm}[H]
	\label{algo:3}
	\SetAlgoLined
	\begin{enumerate}
        \item [1.] Generate $B$ bootstrap samples of size $n$ from the original sample with replacement. Denote the $b$th bootstrap samples as $\{y^{(b)}_i,\mathbf{x}^{(b)}_i\}$, where $b=1,\dots,B$.
	
	\item [2.] For each bootstrap $\{y^{(b)}_i,\mathbf{x}^{(b)}_i\}$, obtain the estimator for the SDR subspace using Algorithm \ref{algo:1} and denote it as $\widehat{\mathcal{S}}^{(b)}(\sigma^2_{u,\ell})$.  
	
	\item [3.] Calculate the distance between $\widehat{\mathcal{S}}^{(b)}(\sigma^2_{u,\ell})$ and $\widehat{\mathcal{S}}(\sigma^2_{u,\ell})$, where $\widehat{\mathcal{S}}(\sigma^2_{u,\ell})$ represents the estimated SDR subspace for the original sample.    Denote this distance as   $D^{(b)}(\sigma^2_{u,\ell})$.
	
	\item [4.] Repeat steps 2 and 3 for each bootstrap sample. Then, calculate the mean distance between $\widehat{\mathcal{S}}^{(b)}(\sigma^2_{u,\ell})$ and $\widehat{\mathcal{S}}(\sigma^2_{u,\ell})$ over the bootstrap samples $b=1,\dots,B$. That is, 
	\begin{equation}
		\overline{D}(\sigma^2_{u,\ell})=\frac{1}{B}\sum_{b=1}^{B}D^{(b)}(\sigma^2_{u,\ell}).
	\end{equation}  
	\end{enumerate}
	\caption{Estimation of the tuning parameter $\sigma_u^2$}
\end{algorithm}
Apply Algorithm \ref{algo:3} for each $\sigma_{u,\ell}^2$, $\ell=1,\dots, k$ to obtain the sequence of average distances $\{\overline{D}(\sigma^2_{u,\ell})\}_{\ell=1}^k$.The optimal value of $\sigma_u^2$ is chosen to be the $\sigma^2_{u,\ell}$ that minimizes $\overline{D}(\sigma^2_{u,\ell})$.
 
\section{R Functions for Integral Transformation Methods}
In this section, we illustrate the functions available in the \textbf{itdr} package that can be utilized for estimating the model parameters and  SDR subspaces using the ITM method. Furthermore, we demonstrate the usage of the functions included in our \textbf{itdr} R package on the {\em automobile} dataset which is accessible within \textbf{itdr}. For more detailed information about the dataset, please visit: \url{https://archive.ics.uci.edu/ml/datasets/automobile}. The dataset consists of $205$ observations for $26$ variables. The response variable is the logarithm of Price,  denoted as $auto\_y=\log(price)$, and the predictor variables are 
represented as ${auto}\_x=(X_1,\cdots, X_{13})$; where $(X_1)$ represents Wheelbase, $(X_2)$ corresponds to Length, $(X_3)$ denotes Width, $(X_4)$ signifies Height, $(X_5)$ indicates Curb Weight, $(X_6)$ refers to Engine Size, $(X_7)$ denotes Bore, $(X_8)$ corresponds to Stroke, $(X_9)$ represents Compression ratio, $(X_{10})$  indicates Horsepower, $(X_{11})$ represents Peak rpm, $(X_{12})$ signifies Cite mpg, and $(X_{13})$ represents Highway mpg. All predictors have been standardized, and missing observations have been excluded.

\subsection{R Function to Estimate the Dimension of the SDR Subspaces}
The {\em d.boots()} function in \textbf{itdr} provides the bootstrap estimator for the dimension of the SDR subspaces, i.e., the dimension of the CS or the CMS.


\begin{example}
    #Install package
    intall.packages("itdr")
    library(itdr)
    data(automobile)
    automobile.na=na.omit(automobile)
    #prepare response and predictor variables
    auto_y=log(automobile.na[,26])
    auto_xx=automobile.na[,c(10,11,12,13,14,17,19,20,21,22,23,24,25)]
    auto_x=scale(auto_xx) #Standardize the predictors
    #call to the d.boots() function with required #arguments
    d_est=d.boots(auto_y,auto_x,Plot=TRUE,space="pdf",
    xdensity = "normal",method="FM")
    auto_d=d_est$d.hat 
\end{example}
The {\em d.boots()} function has the following arguments: {\em $\mathbf{Y}$} is a vector of $n$ observations on the dependent variable; {\em$\mathbf{X}$} is the predictor matrix of dimension $n \times p$; {\em plot} is a logical argument (default is  TRUE); indicating whether to generate the variability plot;  {\em space} is a choice between   ''{pdf}`` for the CS and ''{mean}`` for the CMS;  {\em xdensity} represents the density estimation method for the predictors, which can be ''normal``, ''elliptic``, or ''kernel``; and {\em method} is the transformation method, takes either ''{FM}`` for  Fourier transformation method or ''{CM}`` for convolution transformation.  
\begin{figure}[h!]
\centering
\includegraphics[width=8 cm, height=8 cm]{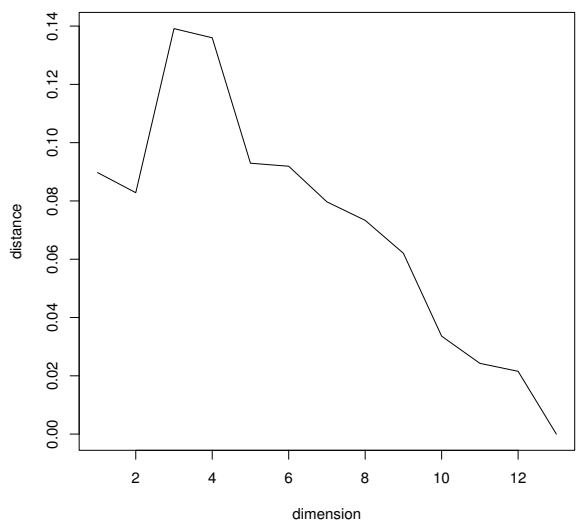}
\caption{Dimension variability plot of {\em automobile dataset}}\label{fig:1}
\end{figure}

Figure \ref{fig:1} depicts the dimension variability plot for the {\em automobile} dataset. Based on  the subspace dimension selection method described in the preceding section, the peak of the overall pattern is identified as $d^{\star}=4$, and the valley point is determined to be  $d_0=2$. Therefore, the estimated dimension of the CS for the {\em automobile} dataset is two, i.e., $\widehat{d}=2$. 
    	
\subsection{R Functions for Estimating the Tuning Parameters}
In this section, we illustrate the usage of the functions provided in the \textbf{itdr}  package to estimate the model parameters. As described in the previous section, we employ the bootstrap procedure to estimate the   tuning parameters of the model. 
\begin{itemize}
\item [1.] Estimation of the predictor's tuning parameter

  The \textbf{itdr} package includes the {\em wx() } function, which is utilized to estimates the  tuning parameter for predictors, denoted as $\sigma^2_{u}$. The following R code illustrates the implementation of the {\em wx()} function to estimate the tuning parameter $\sigma_u^2$.

\begin{example}
    auto_d=2 #The estimated value from d.boots() function
    auto_sw2=wx(auto_y,auto_x,auto_d,wx_seq=seq(0.05,1,by=0.01), wh=1.0, B=50,
    space="pdf",xdensity="normal",method="FM")
    auto_sw2$wx.hat #Estimated Value can be change with the iteration. 
\end{example}

The {\em wx()} function has the following arguments: {\em $\mathbf{Y}$}, a vector of observations of length $n$; {\em $\mathbf{X}$}, the predictor matrix of dimension $n \times p$; {\em $d$}, the estimated dimension of the SDR subspace obtained from the {\em d.boost()} function; {\em wx\_seq}, the candidate list for $\sigma_u^2$; {\em wh}, a fixed value for $\sigma_v^2$ (default $1$); {\em B}, the number of bootstrap samples; the   {\em space} argument with two options: ''{pdf}`` for the CS and ''{mean}`` for the CMS; the   {\em xdensity} argument with three options: ''{normal}`` (the default) for the normal density, ''{kernel}`` for the kernel smoother, and ''{elliptic}`` for the elliptically contoured distribution; and the   {\em method} argument with ''{FM}`` for the Fourier transformation method and ''{CM}`` for the convolution transformation method. The output is the optimal value for $\sigma_u^2$, which for the {\em automobile} dataset yields $\sigma_u^2=0.09$.      
		
\item [2.] Estimation of the response’s tuning parameter

The {\em wy()} function estimates the response tuning parameter,  $\sigma_v^2$. The following R code illustrates the usage of  the {\em wy()} function to tune the parameter
 $\sigma_v^2$.  
\begin{example}
    auto_d=2 # Estimated value from d.boots() function
    set.seed(107) 
    auto_st2=wy(auto_y,auto_x,auto_d,wx=0.14,wy_seq=seq(0.1,1,by=0.1), B=50,
    xdensity="normal",method="FM")
    auto_st2$wy.hat #Estimated Value can be change with the iteration.
\end{example}

The {\em wy()} function has the following arguments: {\em $\mathbf{Y}$} is a vector of $n$ observations on the dependent variable; {\em $\mathbf{X}$} is the predictor matrix of dimension $n \times p$; {\em  $d$} is the estimated dimension of the SDR subspace obtained from the {\em d.boost()} function; {\em wx} (default value is $0.1$) is the estimated value from the {\em wx()} function; {\em wy\_seq} is the candidate list for $\sigma_v^2$; {\em B} is the number of bootstrap samples; the argument {\em xdensity} takes three options: ''{normal}`` (the default) for the normal density, ''{kernel}`` for the kernel smoother, and ''{elliptic}`` for the elliptical contoured distribution; and the argument {\em method} takes ''{FM}`` for the Fourier transformation method and ''{CM}`` for the convolution transformation method. The output is the optimal value for $\sigma_v^2$, which returns $\sigma_v^2=0.9$ for the {\em automobile} dataset.  
			
\item [3. ] Kernel density bandwidth estimation 

When the {\em xdensity} argument is set to ''{\em kernel}``, it is necessary to determine the bandwidth parameter of the kernel density smoother, i.e., $h$. The following R code illustrates the utilization of the {\em wh()} function for estimating the bandwidth parameter $h$.

\begin{example}
    set.seed(109)
    h_hat=wh(auto_y,auto_x,auto_d,wx=0.14,wy=0.9,wh_seq=seq(0.1,2,by=.1),B=50,
    space = "pdf",method="FM")
    #Bandwidth estimator for Gaussian kernel density estimation for CS
    h_hat$h.hat #Estimated Value can be change with the iteration.
\end{example}

The {\em wh()} function has the following arguments: {\em $\mathbf{Y}$}, a vector of observations with length $n$; {\em $\mathbf{X}$}, the predictor matrix of dimension $n \times p$; {\em $d$}, the estimated dimension of the SDR subspace obtained from the {\em d.boost()} function; {\em wx}, the estimated value for $\sigma_u^2$ from {\em wx()} function; {\em wy} is the estimated value for $\sigma_v^2$ obtained from the {\em wy()} function; {\em wh\_seq}, the candidate list for $h$; {\em B}, the number of bootstrap samples; the  {\em space} argument with two options: ''{pdf}`` for the CS and ''{mean}`` for the CMS; and the   {\em method} argument  with ''{FM}`` for the Fourier transformation method and ''{CM}`` for the convolution transformation method. The output of the {\em wh} is the estimated value of the bandwidth parameter $h$, which is determined to be $h=0.1$ for the {\em automobile} dataset. 

\end{itemize}
	
\subsection{R Function for Estimating  Candidate Matrices}

  This section explains the usage of the {\em itdr()} function in our \textbf{itdr} package to estimate the candidate matrices, namely $\mathbf{M}_{ITM}$ or $\mathbf{M}_{ITC}$,  introduced in the previous section. To use the function, we assume the dimension of the CMS (or the CS) is known, which can be estimated using the {\em d.boots()} function. Furthermore,  the following assumptions are made: (i) if the CMS is to be estimated, the parameter $\sigma_u^2$ is tuned; (ii) if the CS is to be estimated, both parameters $\sigma_u^2$ and $\sigma_v^2$ are tuned; (iii) if a kernel density smoother is employed  to estimate the predictor's density, the parameter $h$ is tuned.   The following R code illustrates the application of the {\em itdr()} function to estimate the CS in the {\em automobile} dataset. 
 
\begin{example}
    wx=.14; wy=.9; d=2;
    set.seed(109)
    fit.F_CMS=itdr(auto_y,auto_x,d,wx,wy,space="pdf",xdensity = "normal",method="FM")
    round(fit.F_CMS$eta_hat,2)	
    newx = auto_x 
    plot(auto_y ~ newx[,1], xlab = "First reduced predictor", 
    ylab = paste0(expression(log),'(price)', sep="") )
    plot(auto_y ~ newx[,2], xlab = "Second reduced predictor", 
    ylab = paste0(expression(log),'(price)', sep="") )   
\end{example} 
\begin{singlespace}
\begin{example}
           [,1]  [,2]
     [1,] -0.09  0.01
     [2,]  0.38 -0.16
     [3,] -0.08  0.05
     [4,] -0.11  0.03
     [5,] -0.70 -0.24
     [6,]  0.06  0.83
     [7,]  0.07 -0.14
     [8,]  0.18 -0.13
     [9,] -0.17 -0.08
    [10,] -0.43 -0.26
    [11,] -0.04  0.06
    [12,] -0.26  0.29
    [13,]  0.09 -0.15
\end{example}
\end{singlespace}
The {\em itdr()} function accepts the following arguments: {\em $\mathbf{Y}$}, a vector of $n$ observations on the dependent variable; {\em $\mathbf{X}$}, the predictor matrix of dimension $n \times p$; {\em $d$}, the estimated dimension of the SDR subspace obtained from the {\em d.boots()} function; {\em wx}, the estimated value of $\sigma_u^2$; {\em wy}, the estimated value of $\sigma_w^2$; {\em space}, which can be either ''{pdf}`` for the CS or ''{mean}`` for the CMS; {\em xdensity}, which can be   ''{\em normal}``, ''{\em elliptic}``, or ''{\em kernel}``; and the   {\em method} argument, which can be  either ''{FM}`` for the Fourier transformation method or ''{CM}`` for the convolution transformation method.

\begin{figure}
\centering
\subfigure{\includegraphics[width=0.45\textwidth]{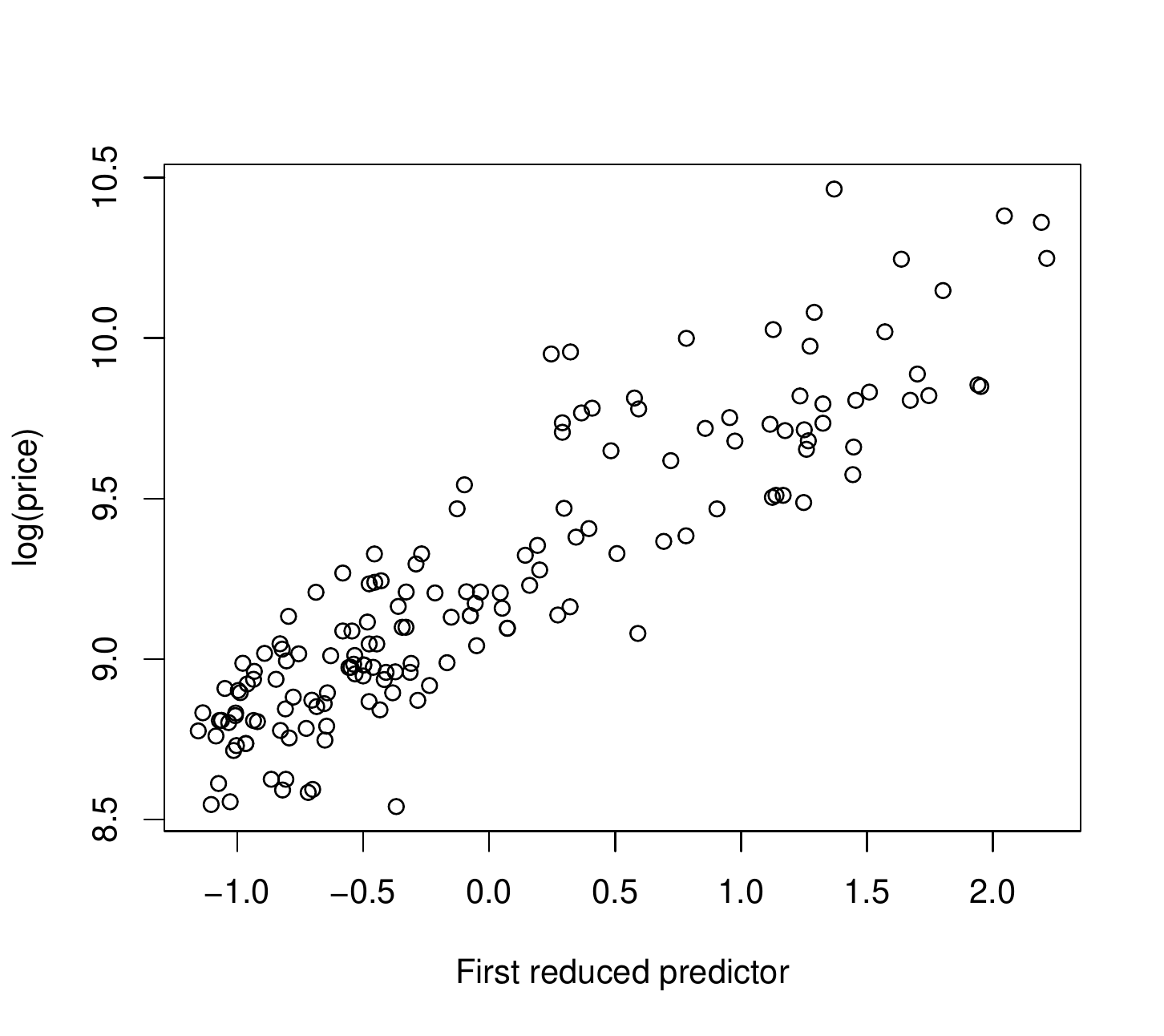}}
\centering
\subfigure{\includegraphics[width=0.45\textwidth]{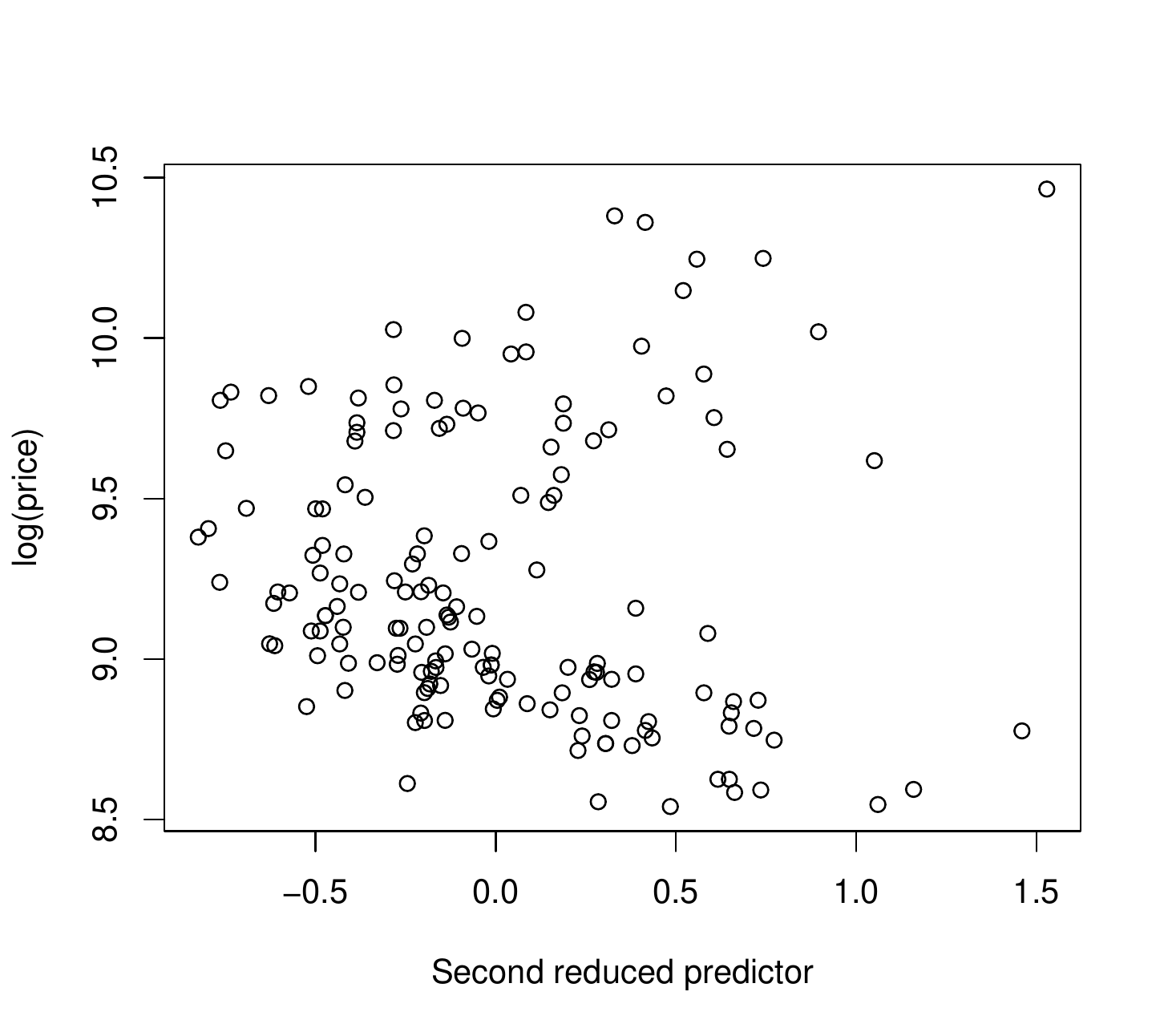}}
\caption{The scatter plots of $\log$(price) versus the first two reduced predictors}
\label{fig:2}
\end{figure}

The columns $(\hat{\beta}_1, \hat{\beta}_2) \in \mathbb{R}^{13 \times 2}$ above represent the basis vectors of the CS for the {\em automobile} dataset. In Figure \ref{fig:2}, the first reduced predictor, $\hat{\beta}_1^T\mathbf{X}$, captures the linear pattern of the response $log$(price), while the second reduced predictor, $\hat{\beta}_2^T\mathbf{X}$, exhibits a nonlinear relationship with the response variable. 

\section{Iterative Hessian Transformation Method for Estimating the Central Mean Subspace (CMS)}

In this section, we provide a summary of the main features of the iterative Hessian transformation (IHT) method proposed by \cite{CL02} for estimating the central mean subspace (CMS) in regression analysis.

Let $\{Y_i,\mathbf{X}_i\}_{i=1}^n$ be an iid sample of size $n$ from the variables $(Y,\mathbf{X})$, where  $\mathbf{X}$ is a $p$-dimensional predictor vector and $Y$ denotes the univariate response variable. We define $\mathbf{Z}=\boldsymbol\Sigma^{-1/2}_{xx}(\mathbf{X}-E[\mathbf{X}])$, where $\boldsymbol\Sigma_{xx}=\text{cov}(\mathbf{X})$ is assumed to be positive definite. The relationship between the central mean subspace (CMS) obtained based on the original predictors and the standardized predictors is given by
$\mathcal{S}_{E[Y\vert\mathbf{X}]}=\boldsymbol\Sigma^{-1/2}_{xx}\mathcal{S}_{E[Y \vert \mathbf{Z}]}$ \cite[][]{CL02}.
 The following lemma, known as Theorem 3 in \cite{CL02}, forms the foundation of the IHT method. 
     
\begin{Lemma}\label{th:3}
	Suppose $E[\mathbf{Z}\vert \boldsymbol\eta^T\mathbf{Z}]$ is linear in $\mathbf{Z}$ and $U$ and $V$ are measurable functions of $\boldsymbol\eta^T\mathbf{Z}$ such that $(UY+V)\mathbf{Z}$ is integrable. Then, $E\{(UY+V)\mathbf{Z}\} \in \mathcal{S}_{E[Y\vert \mathbf{Z}]}$.
\end{Lemma}

This lemma presents an approach to generate basis vectors for constructing the CMS. These vectors, known as COZY vectors,  capture the covariance between $\mathbf{Z}$ and the transformed response $Y^{\star}=(UY+V)$. 
Further details can be found in \cite{CL02}. 
Setting $U=1$ and $V=0$, we have $E[(UY+V)]\mathbf{Z}]=\boldsymbol\Gamma_{yz}$ with probability 1, as stated in  Lemma \ref{th:3}. Furthermore, under condition C.1 in Theorem 1 in \cite{CL02},  we have $\boldsymbol\Gamma_{yz} \in \mathcal{S}_{E[Y\vert \mathbf{Z}]}$. Hence,   an estimator for the CMS can be derived based on the $\boldsymbol{\Gamma}_{yz}$. Lemma \ref{lem:2.6} \cite[Proposition 3 in][]{CL02} provides the detailed estimation procedure for the CMS, known as the iterative Hessian transformation (IHT) method. The following lemma restates Proposition 3 from \cite{CL02}. 

\begin{Lemma}\label{lem:2.6}
Suppose $\mathbf{M} \in \mathbb{R}^{p \times p}$ is a matrix and $\boldsymbol\Gamma\in \mathbb{R}^p$ is a vector. For any $s>r$, if  $~~\mathbf{M}^r\boldsymbol\Gamma \subset Span(\boldsymbol\Gamma,\mathbf{M}\boldsymbol\Gamma,\cdots, \mathbf{M}^{r-1}\boldsymbol\Gamma)$, then, we have  $\mathbf{M}^s\boldsymbol\Gamma \subset Span(\boldsymbol\Gamma,\mathbf{M}\boldsymbol\Gamma,\cdots, \mathbf{M}^{r-1}\boldsymbol\Gamma)$.
\end{Lemma} 
Suppose  $\boldsymbol\Sigma_{yzz}=E[Y\mathbf{Z}\mathbf{Z}^T]$. According to  Corollary 2 in \cite{CL02}, we have $Span\{\boldsymbol\Sigma^j_{yzz}\boldsymbol\Gamma_{yz};~j=0,1,\dots\} \in \mathcal{S}_{E[Y\vert \mathbf{Z}]}$, which has dimension $d$.   By applying  Lemma \ref{lem:2.6},  we can find an integer $s \leq d$ such that the first $s$ vectors in the sequence, $\boldsymbol\Gamma_{yz},\boldsymbol\Sigma_{yzz}\boldsymbol\Gamma_{yz},\dots, \boldsymbol\Sigma_{yzz}^{s-1}\boldsymbol\Gamma_{yz}$, are linearly independent, and all the remaining vectors are linearly dependent on the sequence $\boldsymbol\Gamma_{yz},\boldsymbol\Sigma_{yzz}\boldsymbol\Gamma_{yz},\dots, \boldsymbol\Sigma_{yzz}^{s-1}\boldsymbol\Gamma_{yz}$.  To compute the first $p$   COZY vectors, we  set $U=1$ and $V=0$.  Each COZY vector is then assigned  as a column in the matrix   $\mathbf{M}=(\boldsymbol\Gamma_{yz},\boldsymbol\Sigma_{yzz}\boldsymbol\Gamma_{yz},\dots, \boldsymbol\Sigma_{yzz}^{p-1}\boldsymbol\Gamma_{yz})$, which forms   a $p \times p$ matrix. Then, we define the matrix $\boldsymbol{\Psi}=\mathbf{MM}^T$, and perform the eigenvector and eigenvalue decomposition on $\boldsymbol{\Psi}$. The leading $d$  eigenvectors of $\boldsymbol{\Psi}$,  corresponding to the $d$ largest eigenvalues, serve as the estimated basis for the CMS, $\mathcal{S}_{E[y\vert \mathbf{z}]}$. 

Consider an iid sample $\{Y_i,\mathbf{X}_i\}_{i=1}^n$ and let $\mathbf{Z}$ represent the standardized predictors. Let $\widehat{\mathbf{M}}$ be  the sample version of $\mathbf{M}=(\boldsymbol\Gamma_{yz},\boldsymbol\Sigma_{yzz}\boldsymbol\Gamma_{yz},\dots, \boldsymbol\Sigma_{yzz}^{p-1}\boldsymbol\Gamma_{yz})$. Then, the following steps  outline the estimation process of the CMS using the IHT method \cite[][]{CL02}.

\begin{algorithm}[H]
\label{algo:8}
\SetAlgoLined
\begin{enumerate}
    \item [1.]   Compute the $p$-dimensional COZY vectors: $\widehat{\boldsymbol\Gamma}_{{y}{z}},\widehat{\boldsymbol\Sigma}_{{y}{z}{z}} \widehat{\boldsymbol\Gamma}_{{y}{z}},\dots,\widehat{\boldsymbol\Sigma}_{{y}{z}{z}}^{p-1} \widehat{\boldsymbol\Gamma}_{{y}{z}}$ using the following formulas
	\begin{align}
		\begin{split}
		\widehat{\boldsymbol\Sigma}_{yzz}&=\frac{1}{n}\sum_{i=1}^ny_i\mathbf{ZZ}^T,~~~~
		\widehat{\boldsymbol\Gamma}_{yz}=\frac{1}{n}\sum_{i=1}^ny_i\mathbf{Z},
		\end{split}
	\end{align} 
where $\mathbf{Z}=\widehat{\boldsymbol\Sigma}_{xx}^{-1/2}(\mathbf{X}-\overline{\mathbf{X}})$, and $\widehat{\boldsymbol\Sigma}_{xx}$ and $\overline{\mathbf{X}}$ are the sample covariance matrix and the sample mean of $\mathbf{X}$, respectively. 
	\item [2.] Calculate the $p \times p$ matrix $\widehat{\mathbf{M}}=(\widehat{\Gamma}_{{y}{z}},\widehat{\boldsymbol\Sigma}_{{y}{z}{z}} \widehat{\Gamma}_{{y}{z}},\cdots,\widehat{\boldsymbol\Sigma}_{{y}{z}{z}}^{p-1} \widehat{\Gamma}_{{y}{z}})$, and  compute $\widehat{\boldsymbol\Psi}=\widehat{\mathbf{M}}\widehat{\mathbf{M}}^T$. 
	\item [3.] Perform a spectral decomposition of matrix $\widehat{\boldsymbol\Psi}$. Let $(\widehat{\mathbf{e}}_1,\widehat{\lambda}_1), \cdots, (\widehat{\mathbf{e}}_d,\widehat{\lambda}_d)$ be the first $d$ leading eigenvectors-eigenvalue pairs, where $\widehat{\lambda}_1\geq\cdots \geq \widehat{\lambda}_d\geq 0=\cdots=0.$ 
	\item [4.] The estimated CMS is given by $\widehat{\mathcal{S}}_{E[Y\vert \mathbf{X}]}=Span\{\widehat{\boldsymbol\Sigma}_{xx}^{-1/2}\widehat{\mathbf{e}}_1,\dots,\widehat{\boldsymbol\Sigma}_{xx}^{-1/2}\widehat{\mathbf{e}}_d \}$ where $\widehat{\boldsymbol{\Sigma}}_{xx}$ is the sample covariance matrix. 
\end{enumerate}
\caption{IHT method for Estimating the CMS}
\end{algorithm}

\subsection{R Function for Iterative Hessian Transformation Method}

The {\em itdr()} function in our \textbf{itdr} package facilitates the estimation of the central mean subspace by using the IHT method. The following R code demonstrates the process of estimating the CMS on the {\em Recumbent cows} dataset, which was collected at the Ruakura (N. Z.) and analyzed by \cite{Clark87}. We have used the same response variables and predictor variables as \cite{CL02}. Specifically, the predictors are $\log(AST)$ (logarithm of serum aspartate aminotransferase in U/l at 30C); $log(CK)$ (logarithm of serum creatine phosphokinase in U/l
at 30C); and $\log(UREA)$ (logarithm of serum urea in mmol/l). The response variable is a binary variable, where $Y = 1$ indicates surviving cows, and $Y = 0$ indicates not surviving cows.
\begin{example}
    library(itdr)
    data("Recumbent")
    Recumbent.df=na.omit(Recumbent)
    y=Recumbent.df$outcome
    X1=log(Recumbent.df$ast)
    X2=log(Recumbent.df$ck)
    X3=log(Recumbent.df$urea)
    x=matrix(c(X1,X2,X3),ncol=3)
    fit.iht_CMS=itdr(y,x,d = 2,method="iht")
    fit.iht_CMS$eta_hat
\end{example}
The {\em itdr()} function has the following arguments for the IHT method: {\em $\mathbf{Y}$} is a vector of $n$ observations on the response variable, {\em $\mathbf{X}$} is the predictor matrix of dimension $n\times p$, {\em $d$} is the dimension of the central mean subspace, and the {\em method} argument takes ``iht'' to indicate the use of the IHT method.  

\begin{singlespace}
  \begin{example}
               [,1]        [,2]
    [1,]  0.3260269  0.95986216
    [2,] -0.2395713 -0.27884564
    [3,] -0.9145010  0.03016189
\end{example}  
\end{singlespace}

The output displayed above illustrates the estimated basis vectors for the CMS of the {\em Recumbent cows} dataset.  

\section{Fourier Transformation Method for Inverse Dimension Reduction in Multivariate Regression}

In this section, we present a summary of the theoretical framework of the Fourier transformation method for inverse dimension reduction (invFM) proposed by \cite{Weng18}. This method aims to estimate the central subspace (CS). We consider an iid sample $\{\mathbf{Y}_i,\mathbf{X}_i\}$, $i=1,\dots,n$, from $(\mathbf{Y},\mathbf{X})$, where $\mathbf{Y} \in \mathbb{R}^{q}$ represents a $q$-dimensional response variable, $\mathbf{X} \in \mathbb{R}^p$ is a $p$-dimensional predictor variable, and $n$ is the sample size. Furthermore, we assume that $\mathbf{Z}$ represents the standardized version of the predictor $\mathbf{X}$, defined as $\mathbf{Z}=\boldsymbol\Sigma^{-1/2}(\mathbf{X}-\boldsymbol\mu)$, where $\boldsymbol\mu$ and $\boldsymbol\Sigma$ are the mean and covariance matrix of $\mathbf{X}$, respectively. Under the linearity condition, it can be shown that  $m(\mathbf{y})=E[\mathbf{Z}\vert \mathbf{Y}=\mathbf{y}]\in \mathcal{S}_{\mathbf{Y}\vert \mathbf{Z}}$ \cite[][]{cook98}. Let $f(\mathbf{y})$ denote the marginal distribution of $\mathbf{Y}$ and let $\boldsymbol\omega \in \mathbb{R}^{q}$. According to \cite{Weng18}, the Fourier transformation of the density-weighted $m(\mathbf{y})$ is given by
\begin{equation}\label{eq:25}
	\boldsymbol\psi(\boldsymbol\omega)=\int \exp\{i\boldsymbol\omega^T\mathbf{y}\}m(\mathbf{y})f(\mathbf{y})d\mathbf{y}.
\end{equation}  
For more comprehensive information regarding the Fourier transformation, see \cite{Folland92}.  Using simple algebra, we can show that $\boldsymbol\psi(\boldsymbol\omega)=E[\exp\{i\boldsymbol\omega^T\mathbf{y}\}\mathbf{Z}]$, and $\mathcal{S}_{E[\mathbf{Z}\vert \mathbf{Y}]}=Span\left\{\boldsymbol\psi(\boldsymbol\omega); \boldsymbol\omega\in\mathbb{R}^q\right\}$. Moreover, under the linearity condition,  $\mathcal{S}_{E[\mathbf{Z}\vert \mathbf{Y}]}$ can be expressed as  
$ Span\left\{m(\mathbf{y}), \mathbf{y} \in supp(f(\mathbf{y}))\right\}=Span\left\{m(\mathbf{y})f(\mathbf{y}), \mathbf{y}\in supp(f(\mathbf{y}))\right\}\subseteq \mathcal{S}_{\mathbf{Y}\vert \mathbf{Z}}$. Consequently, by employing the inverse Fourier transformation provided in \cite{Folland92}, we can recover the density-weighted conditional mean function $m(\mathbf{y})$ from $\psi(\boldsymbol\omega)$, as follows
\begin{equation}\label{eq:18.1}
	m(\mathbf{y})f(\mathbf{y})=\frac{1}{2\pi}\int \exp\{-i\boldsymbol\omega^T\mathbf{y}\}\boldsymbol\psi(\boldsymbol\omega)d\boldsymbol\omega.
\end{equation}
Hence, we have $\mathcal{S}_{E[\mathbf{Z}\vert\mathbf{Y}]}=Span\{\boldsymbol\psi(\boldsymbol\omega); \boldsymbol\omega \in \mathbb{R}^{q}\}=Span\{a(\boldsymbol\omega),b(\boldsymbol\omega); \boldsymbol\omega\in \mathbb{R}^{q}\}$ where $\mathbf{a}(\boldsymbol\omega)$ and $\mathbf{b}(\boldsymbol\omega)$ are the real and imaginary parts of $\boldsymbol\psi(\boldsymbol\omega)$, respectively. 

 According to the preceding discussion, the Fourier transform expression $\boldsymbol\psi(\boldsymbol\omega)$ does not explicitly contain the mean function $m(\mathbf{y})$. Consequently, $\boldsymbol\psi(\boldsymbol\omega)$ can be estimated without  directly estimating $m(\mathbf{y})$. That is, we can find an estimator for the CS by calculating $\boldsymbol\psi(\boldsymbol\omega)$ for a given set of $q$-dimensional vectors $\{\boldsymbol\omega_r\}_{r=1}^k$, where $k$ is a prespecified number. Based on Proposition 2.1 in \cite{Weng18}, there exists a finite sequence of $\boldsymbol\omega_r \in \mathbb{R}^q$, $r=1,\dots,k$, such that $\mathcal{S}_{E[\mathbf{Z}\vert \mathbf{Y}]}=Span\{\mathbf{a}(\boldsymbol\omega_1),\mathbf{b}(\boldsymbol\omega_1),\dots,\mathbf{a}(\boldsymbol\omega_k),\mathbf{b}(\boldsymbol\omega_k)\}$.
Therefore, we utilize $\boldsymbol\psi(\boldsymbol\omega_r)$s to derive a candidate matrix that targets the CS. We construct a nonnegative definite symmetric dimension reduction matrix  $\mathbf{V}\in \mathbb{R}^{p \times p}$  using $\boldsymbol\psi(\boldsymbol\omega_r)$, $r=1,\dots,k$, which is referred to as a kernel dimension reduction matrix. Then, we perform the eigen decomposition of a sample version of $\mathbf{V}$, denoted as  $\widehat{\mathbf{V}}$. Finally, the 
first $d$ leading eigenvectors corresponding to the foremost $d$ leading eigenvalues of $\widehat{\mathbf{V}}$ form an orthogonal basis for the CS. 

\subsection{Algorithm of the invFM}
Let $\{\boldsymbol\omega_r\}_{r=1}^k$ be a predefined set of $q$-dimensional vectors, $(\mathbf{Y}_i,\mathbf{X}_i)$, $i=1,\dots,n$, be a random sample of size $n$, where $\mathbf{Y}_i \in \mathbb{R}^q$, and $\mathbf{X}_i \in \mathbb{R}^p$. Suppose $\boldsymbol\Omega=(\mathbf{a}(\boldsymbol\omega_1),\mathbf{b}(\boldsymbol\omega_1),\cdots,\mathbf{a}(\boldsymbol\omega_k),\mathbf{b}(\boldsymbol\omega_k))$, for $k>0$, and let $\mathbf{V}=\boldsymbol\Omega\boldsymbol\Omega^T$ represent the population version of the  kernel dimension reduction matrix, and assume the dimension $d=dim(\mathcal{S}_{E(\mathbf{Z}\vert \mathbf{Y})})$ is known.  Then, the following standard steps are used to estimate the central subspace based on the Fourier transformation method for inverse dimension reduction.    

\begin{algorithm}[H]
	\label{algo:4}
	\SetAlgoLined
	\begin{enumerate}
        \item [1.] Standardize the predictors  by computing  $\widehat{\mathbf{z}}_i= \widehat{\boldsymbol{\Sigma}}^{-1/2}(\mathbf{x}_i-\overline{\mathbf{x}}) for each observation i=1,\dots, n$, where $\overline{\mathbf{x}}$ and $\widehat{\boldsymbol{\Sigma}}$ are the sample mean and the sample covariance matrix of $\mathbf{X}$, respectively. 
	\item [2.] Choose a  random sequence of  $\{\boldsymbol{\omega}_r\}_{r=1}^k$ from a normal distribution and compute the sample version of $\boldsymbol{\psi}(\boldsymbol{\omega}_r)$ using the formula:
	\begin{equation}
	 \widehat{\boldsymbol{\psi}}(\boldsymbol{\boldsymbol\omega_r})=n^{-1}\sum_{i=1}^{n}\exp\{i\boldsymbol{\boldsymbol\omega_r^T\mathbf{y}_i}\}\widehat{\mathbf{z}}_i,
	\end{equation}
	 calculate the real and imaginary parts as $\widehat{\mathbf{a}}(\boldsymbol{\omega_r})=Real \{\widehat{\boldsymbol{\psi}}(\boldsymbol{\omega}_r)\}$ and $\widehat{\mathbf{b}}(\boldsymbol{\omega}_r)=Imag\{\widehat{\boldsymbol{\psi}}(\boldsymbol{\omega_r})\}$.
	\item [3.] Construct $\widehat{\boldsymbol{\Omega}}$ and $\widehat{\mathbf{V}}$ as follows
	\begin{equation}\label{eq:28}
		\widehat{\boldsymbol{\Omega}}=\{\widehat{\boldsymbol{a}}(\boldsymbol{\omega}_r),\widehat{\boldsymbol{b}}(\boldsymbol{\omega}_r)\}_{r=1}^k,~~~ \widehat{\mathbf{V}}= \widehat{\boldsymbol{\Omega}}\widehat{\boldsymbol{\Omega}}^T,
	\end{equation}
	where $\widehat{\boldsymbol{\Omega}}$ is a $p\times2k$ matrix and $\widehat{\mathbf{V}}$ is a $p \times p$ sample kernel dimension reduction matrix.
	\item [4.] Perform eigen decomposition of $\widehat{\mathbf{V}}$ and select the first $d$ leading eigenvectors $(\widehat{\boldsymbol{\eta}_i},i=1,\dots,d)$   corresponding to the $d$ largest eigenvalues $\widehat{\lambda}_1\geq \cdots \geq \widehat{\lambda}_d$ as the estimated basis of $\mathcal{S}_{E[\mathbf{Z}\vert \mathbf{Y}]}$. 
	\item [5.] Back transform the eigenvectors to the original scale of $\mathbf{X}$. Compute  $\widehat{\boldsymbol{\beta}}_i=\widehat{\boldsymbol{\Sigma}}^{-1/2}\widehat{\boldsymbol{\eta}}_i,~~i=1,\dots,d$. 
	\end{enumerate}
	\caption{Fourier transformation method }
\end{algorithm}
The above steps are standard in SDR procedures except Step 3, which employs the proposed invFM method. In the case of  a  univariate response,  select scalar values for $\omega_r$, i.e., $\omega_r \in \mathbb{R}$.

\subsection{Dimension Selection of the Central Subspace using invFM Method}
In the previous section, we made the assumption that the dimension $d$ of the central subspace (CS) is already known. In this section, we describe a testing procedure to select the appropriate dimension. Let us consider  a hypothetical dimension $m$  for  the CS, and formulate  the null hypothesis ($H_0$) and alternative hypothesis ($H_a$) as follows:
\begin{equation}\label{eq:hy}
	H_0: d=m~~~~\text{versus}~~~~ H_a: d>m.
\end{equation}
\cite{Weng18} proposed a weighted chi-square test statistic to test the hypothesis in Equation \eqref{eq:hy}. This statistic, denoted as $\widehat{\Lambda}_m$, is defined as
\begin{equation}\label{eq:30}
	\widehat{\Lambda}_m=n\sum_{j=m+1}^{p}\widehat{\lambda}_j.
\end{equation}  
 The value of $m$ begins at $m=0$, and continues to increase by one until the null hypothesis at the current value of $m$ cannot be rejected. Moreover, a scaled test statistic suggested by \cite{Bentler2000}  as a simplified version of the weighted chi-square test, can be used to determine the dimension. The scaled test statistic is defined as
\begin{equation}\label{eq:31}
	\overline{T}_m=[\text{trace}(\widehat{\mathbf{V}})/p^{\star}]^{-1}n\sum_{j=m+1}^{p}\widehat{\lambda}_j \sim \chi^{2}_{p^{\star}},
\end{equation}  
where $\widehat{\mathbf{V}}$ is a consistent estimator of $\mathbf{V}$, and is defined in Equation \eqref{eq:28}, and $p^{\star}=(p-m)(2k-m)$.
Another test statistic that can be employed  to test the hypotheses in Equation \eqref{eq:hy} is the adjusted test statistic proposed by \cite{Bentler2000}, defined as
\begin{equation}\label{eq:32}
	\widetilde{T}_m=[\text{trace}(\widehat{\mathbf{V}})/s^{\star}]^{-1}n\sum_{j=m+1}^{p}\widehat{\lambda}_j \sim \chi^{2}_{s^{\star}},
\end{equation}
where $s^{\star}=[\text{trace}(\widehat{\mathbf{V}})]^2/\text{trace}(\widehat{\mathbf{V}}^2)$, and $\widehat{\mathbf{V}}$ is defined in Equation \eqref{eq:28}.

\section{R Functions for Fourier Method for Inverse Dimension Reduction in Multivariate Regression}

In this section, we demonstrate the usage of functions included in our \textbf{itdr} package to perform the Fourier transformation method for the inverse dimension reduction proposed by \cite{Weng18}.  The first subsection illustrates the process of selecting the dimension of the central subspace (CS) using the ``2015 Planning Database'' ({\em  PDB}) dataset which is available in our \textbf{itdr} package. More details of this dataset can be found at \url{https://www.census.gov/data/datasets/2015/adrm/research/2015-planning-database.html}. The main function for estimating the central subspace is described in the second subsection.
   
\subsection{R Function for Determining the Dimension of the Central Subspace (CS)}

Within our $\mathbf{itdr}$ package, the   {\em d.test()} function calculates the $p$-values for three different test statistics defined in Equations \eqref{eq:30}-\eqref{eq:32}.  These include the weighted chi-square test statistic $\widehat{\Lambda}_m$, the scaled test statistic $\overline{T}_m$, and the adjusted test statistic $\widetilde{T}_m$.     

We utilize the same dataset  employed by \cite{Weng18}, which consists of the 2010 Census and 2009-2013 American Community Survey data,   This dataset encompasses housing, demographic, socioeconomic, and Census operational information  at the block-group level.
Any observations with missing values are excluded. 
Then, the Box-Cox transformation is applied to the predictors to ensure adherence to the linearity condition.
\begin{example}
    library(itdr)
    data(PDB)
    colnames(PDB)=NULL
    p=15
    set.seed(123)
    df=PDB[,c(79,73,77,103,112,115,124,130,132,145,149,151,153,155,167,169)]
    dff=as.matrix(df)
    #remove the NA rows
    planingdb=dff[complete.cases(dff),]
    y=planingdb[,1] #n-dimensionl response vector
    x=planingdb[,c(2:(p+1))] # raw design matrix
    x=x+0.5
    # design matrix after transformations
    xt=cbind(x[,1]^(.33),x[,2]^(.33),x[,3]^(.57),x[,4]^(.33),x[,5]^(.4),
    x[,6]^(.5),x[,7]^(.33),x[,8]^(.16),x[,9]^(.27),x[,10]^(.5),
    x[,11]^(.5),x[,12]^(.33),x[,13]^(.06),x[,14]^(.15),x[,15]^(.1))
    #run the hypothesis tests
    d.test(y,x,m=1)
\end{example}
The {\em d.test()} function takes the following arguments: $\mathbf{Y}$, a vector of $n$ observations; $\mathbf{X}$,  a predictor matrix of dimension $n \times p$; and $m$, the assumed dimension of the central subspace. 
\begin{example}
    	 Hypothesis Tests for selecting sufficient dimension (d)
     Null: d=m	 vs	 Alternative: d>m 
     Test 	 W.Ch.Sq 	Scaled     Adjusted  
     p-value 	 0.9837 	 1       0.9306314
\end{example}
The above output displays the $p$-values for the three different tests introduced in Equations \eqref{eq:30}-\eqref{eq:32}. 
Starting with $m=1$, if the p-value is below a significant level,  the null hypothesis is rejected,  and the test proceeds to the next value of $m$. If the p-value is above the significant level,   the null hypothesis is not rejected, and the test is stopped.
According to the results, all $p$-values from the three different tests are greater than $0.05$.  Therefore, we can assume the hypothetical dimension, i.e., $m=1$, represents the true dimension of the CS in the {\em PDB} dataset, that is, $\widehat{d}=1$. 
	
\subsection{R Function for Estimating the Central Subspace (CS)}

The {\em invFM()} function in our \textbf{itdr} package provides an estimator for the CS using the Fourier transformation method for the inverse dimension reduction proposed by \cite{Weng18}. Similar to the previous section, we utilize the same response variable and the predictor variable with the same Box-Cox transformations as described in \cite{Weng18} on the {\em PDB} dataset. The following R code illustrates the usage of the {\em invFM()} function.
  
\begin{example}
    set.seed(123)
    W=sapply(100,rnorm)
    # estimated dimension of the CS from Section 4.1
    betahat <-invFM(x = xt, y = y, d = 1, w = W, x_scale = F)$beta 
    #estimated basis
    betahat
    plot(y ~ xt 
    ylab = "Health insurance coverage")
\end{example}
\begin{singlespace}
\begin{example}
 [1] -0.14791302 -0.25891444 -0.61782407 -0.10608213 -0.07648674 -0.48468920
 [7] -0.02378407  0.05098216 -0.06333981 -0.35316016 -0.27668059 -0.24190026
[13] -0.09364571 -0.03655110 -0.01670101
\end{example}
\end{singlespace}
The {\em invFM()} function accepts the following arguments: $\mathbf{X}$, the predictor matrix of dimension $n \times p$; $\mathbf{Y}$, the response matrix of dimension $n \times q$; $d$, the dimension of the central subspace; $W$, a prespecified matrix of $\{\boldsymbol\omega_r\}$s of dimension $q \times k$; and {\em x\_scale},  a logical parameter  that specifies whether the predictor matrix should be normalized or not, with the  ''{TRUE}`` as the default. 

\begin{figure}
\centering
{\includegraphics[width=0.5\textwidth]{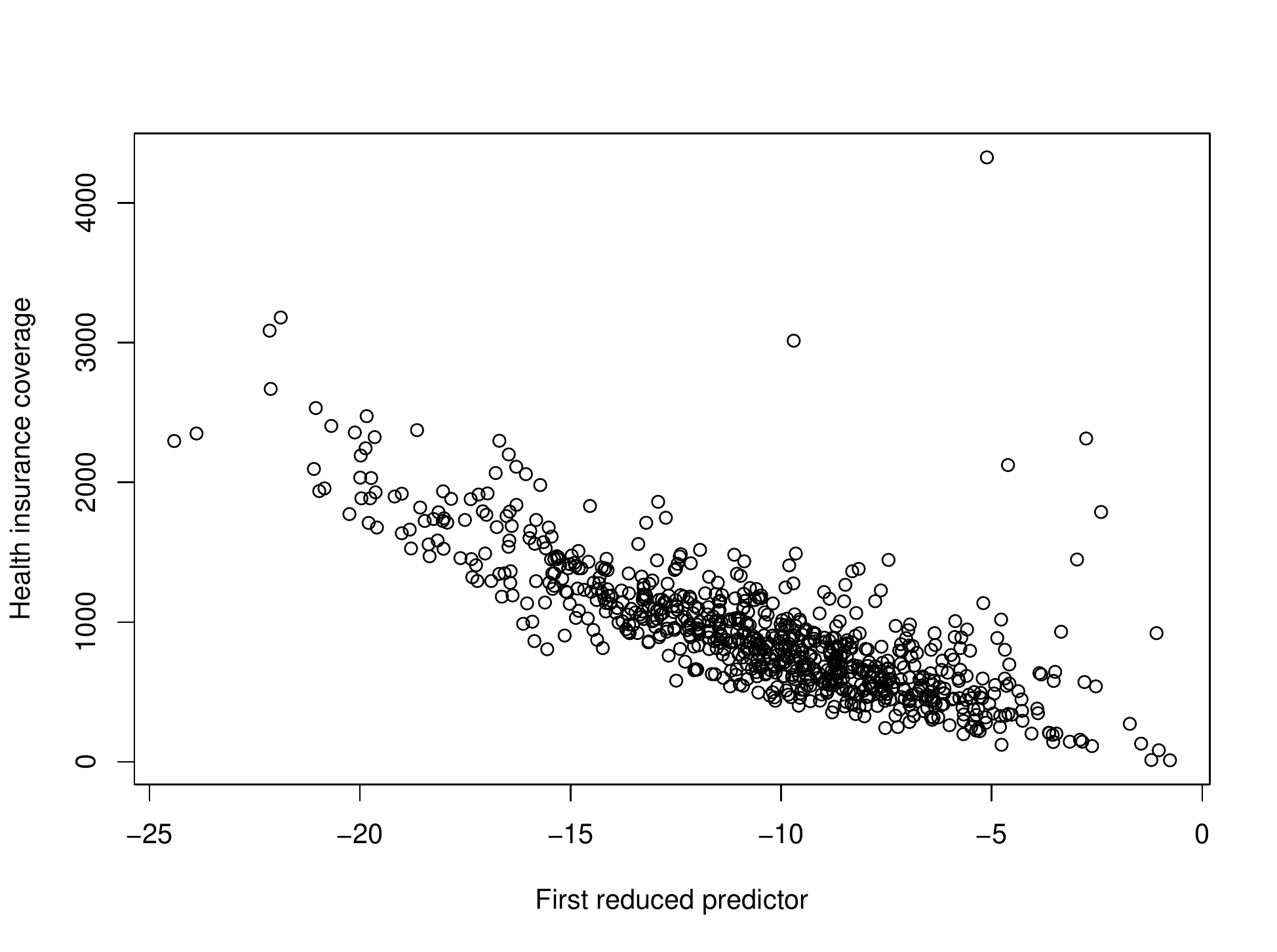} }
\caption{The scatter plot of $Y$: health insurance coverage versus the first reduced predictor}
\label{fig:3}
\end{figure}

The provided output displays the estimated single-index direction $\hat{\boldsymbol{\beta}} \in \mathbb{R}^{15 \times 1}$ for the central subspace of the {\em PDB} dataset. 
Based on the scatter plot presented in Figure \ref{fig:3}, it can be observed that there exists a negative correlation between health insurance coverage and the first reduced predictor. Specifically, as the value of the first reduced predictor increases, the health insurance coverage tends to decrease. This finding suggests that the first reduced predictor holds potential significance as a  predictor variable in the model, and further investigation is warranted.

\section{A Minimum Discrepancy Approach using Fourier Transformation}\label{sec:md}

In this section, we provide an overview of a family of optimal estimators that optimize a quadratic function using the Fourier transformation approach introduced by \citep{weng2022minimum}. Given a finite sequence $\{\boldsymbol\omega_r\} \in \R^q, r = 1, \ldots, k$, we define $\boldsymbol\xi_{r}= \boldsymbol\Sigma^{-1}[E(e^{i\boldsymbol\omega^T_r \mathbf{Y}}\mathbf{X}) - E(e^{i\boldsymbol\omega^T_r \mathbf{Y}})E(\mathbf{X}) ] \in \mathbb{C}^{p}$. 
According to \cite{Weng18}, we have $\boldsymbol\xi_{r} \in \mathcal{S}_{\mathbf{Y}\vert \mathbf{X}}$.
 We denote the real and imaginary parts of a complex vector with superscripts $R$ and $I$, respectively, and we represent $\boldsymbol\xi $ as a matrix $\boldsymbol\xi = \left(\boldsymbol\xi^{R}_{r},\boldsymbol\xi^{I}_{r}\right)^k_{r=1}$, combining each $\boldsymbol\xi^{R}_r$ and $\boldsymbol\xi^{I}_r$. Hence, the column space spanned by $\boldsymbol\xi$ is within the central subspace: $Span(\boldsymbol\xi) \subseteq \mathcal{S}_{\mathbf{Y}\vert \mathbf{X}}$.

Assume $\{\mathbf{y}_j, \mathbf{x}_j\}$, for $j = 1,\ldots,n$, are independent and identically distributed (iid) samples  of  $(\mathbf{Y},\mathbf{X})$. Let $\bar{\mathbf{x}}$  be  the sample mean of $\mathbf{X}$ and 
$$\hat{\boldsymbol\xi}_{r} = \hat{\boldsymbol\Sigma}^{-1}\left(\frac{1}{n}\sum^{n}_{j=1}e^{i\boldsymbol\omega^T_r\mathbf{y}_j}\mathbf{x}_j - \frac{1}{n}\sum^{n}_{j=1}e^{i\boldsymbol\omega^T_r\mathbf{y}_j}\bar{\mathbf{x}}\right)$$ 
denote the sample estimate of $\boldsymbol\xi_{r}$. Then, $ \hat{\boldsymbol\xi} = \left(\hat{\boldsymbol\xi}^{R}_{r},\hat{\boldsymbol\xi}^{I}_{r}\right)^k_{j=r}$ serves as a sample estimate of $\boldsymbol\xi$.
We define the quadratic discrepancy function (QDF) of $\boldsymbol{\Gamma}\in \mathbb{R}^{p\times d}$ and $\mathbf{C}\in \mathbb{R}^{d\times 2k}$, with respect to the inner product matrix $\mathbf{V}$, as:
\begin{equation}\label{eq:quad-FTIRE}
F_d(\boldsymbol{\Gamma}, \mathbf{C};\mathbf{V}) = [\text{vec}(\hat{\boldsymbol\xi}) - \text{vec}(\boldsymbol{\Gamma} \mathbf{C})]^T \mathbf{V} [\text{vec}(\hat{\boldsymbol\xi}) - \text{vec}(\boldsymbol{\Gamma} \mathbf{C})],
\end{equation}
where $\text{vec}(\mathbf{A})$ is the vectorization of a matrix $\mathbf{A}$.
The estimation of the central subspace can be achieved by minimizing the objective function $F_d(\boldsymbol{\Gamma}, \mathbf{C};\mathbf{V})$, where $\boldsymbol{\Gamma}$ represents an orthogonal basis of the central subspace, $\mathbf{C}$ denotes the coordinates of the data points relative to the basis, and $\mathbf{V}$ is an inner product matrix. The accuracy of the resulting estimator heavily depends on the choice of $\mathbf{V}$. To shed light on this issue, \cite{weng2022minimum} conducted an investigation into the effects of five different $\mathbf{V}$ matrices on the estimator and its properties. The final estimate $(\hat{\boldsymbol{\Gamma}},\hat{\mathbf{C}})$ is obtained by minimizing  the objective function and provides an estimate of the central subspace and the corresponding coordinates of the data points.

1. The Fourier transform Inverse Regression Estimator (FT-IRE) is an optimal estimator that achieves asymptotic efficiency without any constraints or strong assumptions. To construct the inner product matrix $\boldsymbol V$, we begin by considering  the population residual   $\epsilon_{r} = e^{i\boldsymbol\omega^T_r \mathbf{Y}} - E e^{i\boldsymbol\omega^T_r \mathbf{Y}} - \mathbf{Z}^{T}E (e^{i\boldsymbol\omega^T_r \mathbf{Y}}\mathbf{Z})$    obtained from an ordinary least squares fit of $e^{i\boldsymbol\omega^T_r \mathbf{Y}}$ on $\mathbf{Z}$.  Furthermore,  $\boldsymbol\epsilon = \left(\epsilon_{1}^{R},\epsilon_{1}^{I},\ldots,\epsilon_{k}^{R},\epsilon_{k}^{I}\right)^T$ consists of both the real and imaginary parts. As a result, we have 
    $$
    \sqrt{n}[\text{vec}(\hat{\boldsymbol\xi})-\text{vec}(\boldsymbol\xi)] \overset{D}{\rightarrow} {N}(\mathbf{0},\boldsymbol\Sigma_{\boldsymbol\xi}),
    $$
    where $\boldsymbol\Sigma_{\boldsymbol\xi} = \text{cov}\{\text{vec}[\boldsymbol\Sigma^{-1/2}\mathbf{Z}\boldsymbol{\epsilon}^{T}]\} \in \mathbb{R}^{2kp\times 2kp}$ is the limiting covariance matrix, and $\overset{D}{\rightarrow}$ denotes the convergence in distribution.  
  When implementing the information matrix $\boldsymbol V = {\boldsymbol\Sigma_{\boldsymbol\xi}}^{-1}$ to the QDF in Equaiton  \eqref{eq:quad-FTIRE}, the resulting minimizer, denoted as  $\hat{\boldsymbol{\Gamma}}$, is the FT-IRE.

 2. When the value of $k$ becomes too large, FT-IRE may encounter a singular limiting covariance matrix. To address this limitation,  
 \citet{weng2022minimum}, proposed a technique where they construct multiple QDFs, each with its own limiting covariance matrix, treating them as independent entities. 
Specifically, they utilized several sequences of $\boldsymbol\omega^{(l)} \in \R^{k_l},$ with $\sum k_l = k, l=1,\ldots, s$ and construct $s$ QDFs with the corresponding limiting covariance matrix $\boldsymbol\Sigma_{\boldsymbol\xi}^{(l)} \in \R^{2pk_l \times 2pk_l}$. 
 The degenerated QDF is defined as the summation of these $s$ QDFs, that is, 
 \begin{equation}\label{eq:FT-DIRE}
 F_d(\boldsymbol{\Gamma}, \mathbf{C};\{\boldsymbol\Sigma_{\boldsymbol\xi}^{(l)}\}) = \sum^{s}_{l=1} [\text{vec} (\hat{\boldsymbol\xi}_l) - \text{vec}(\boldsymbol{\Gamma} \mathbf{C}_l)]^T \boldsymbol\Sigma_{\boldsymbol\xi}^{(l)} [\text{vec} (\hat{\boldsymbol\xi}_l) - \text{vec}(\boldsymbol{\Gamma} \mathbf{C}_l)].
 \end{equation}
 This expression is equivalent to Equation (\ref{eq:quad-FTIRE}) but employs a new inner product matrix $\mathbf{V} = \hat{\boldsymbol{\Gamma}}_D^{-1}= \text{diag}(\{\hat{\boldsymbol\Sigma}_l^{(l)-1} \})$.    The degenerated estimator $\hat{\boldsymbol{\Gamma}}$, which minimizes (\ref{eq:FT-DIRE}), is referred to as the Fourier transform degenerated inverse regression estimator (FT-DIRE).

 3. 
 \cite{weng2022minimum} proved that the invFM estimator is sub-optimal when employing a special inner product matrix $\mathbf{V} = \text{diag}\{\hat{\boldsymbol\Sigma}\}$ in the QDF function \eqref{eq:quad-FTIRE}. The corresponding special QDF can be expressed as follows:
 \begin{equation}\label{eq:FT-SIRE}
 \begin{array}{ccl}
   F_d (\boldsymbol{\Gamma}, \mathbf{C}; \text{diag}\{\hat{\boldsymbol\Sigma}\} ) & = & [\text{vec}(\hat{\boldsymbol\xi}) - \text{vec}( \boldsymbol{\Gamma} \mathbf{C})]^T \text{diag}\{\hat{\boldsymbol\Sigma}\} [\text{vec}(\hat{\boldsymbol\xi}) - \text{vec}( \boldsymbol{\Gamma} \mathbf{C})]\\
   & = & \sum_{l=1}^{2k} (\hat{\boldsymbol\xi}_l - \boldsymbol{\Gamma} \mathbf{C}_l)^T \hat{\boldsymbol\Sigma} (\hat{\boldsymbol\xi}_l - \boldsymbol{\Gamma} \mathbf{C}_l). \\
 \end{array}
 \end{equation}
 In this formulation, each column of $\boldsymbol\xi$ is considered independent of the other. The FT-SIRE is theoretically equivalent to the invFM  estimator proposed by \cite{Weng18}.
The minimizer of \eqref{eq:FT-SIRE}, denoted as Fourier transform special inverse regression estimator (FT-SIRE), is represented by $\hat{\boldsymbol{\Gamma}}$.

4. 
Obtaining a consistent estimate of ${\boldsymbol\Sigma_{\boldsymbol\xi}}^{-1}$ requires considering fourth moments of the predictors. 
To achieve  robust estimation,  let us assume the covariance matrix $\boldsymbol\Sigma$ is known. Let $\tilde{\boldsymbol\xi}_{j} = \boldsymbol\Sigma^{-1}(\frac{1}{n}\sum^{n}_{j=1}e^{i\boldsymbol\omega^T_r \mathbf{y}_j}\mathbf{x}_j - \frac{1}{n}\sum^{n}_{j=1}e^{i\boldsymbol\omega^T_r\mathbf{y}_j}\bar{\mathbf{x}})$, and $ \tilde{\epsilon} = e^{i\boldsymbol\omega^T \mathbf{Y}} - E e^{i\boldsymbol\omega^T \mathbf{Y}}$. Furthermore, we  define $ \tilde{\boldsymbol\xi} = \left(\tilde{\boldsymbol\xi}^{R}_{r},\tilde{\boldsymbol\xi}^{I}_{r}\right)^m_{r=1} \in \mathbb{R}^{p\times 2m}$. Then, we have 
 $$
 \sqrt{n} [\text{vec}(\tilde{\boldsymbol\xi}) - \text{vec}(\boldsymbol\xi)] \overset{D}{\rightarrow} N(0,\tilde{\boldsymbol\Sigma}_{\tilde{\boldsymbol\xi}}),
 $$
 where $\tilde{\boldsymbol\Sigma}_{\tilde{\boldsymbol\xi}} = (\mathbf{I} \otimes \boldsymbol\Sigma^{-1/2}) \text{cov}[\text{vec}(\mathbf{Z}\tilde{\boldsymbol\epsilon}^T)] (\mathbf{I} \otimes \boldsymbol\Sigma^{-1/2})$ and $\tilde{\boldsymbol{\epsilon}} = (\tilde\epsilon_{1}^{R},\tilde\epsilon_{1}^{I},\ldots,\tilde\epsilon_{k}^{R},\tilde\epsilon_{k}^{I})^T$. Here, $\tilde\epsilon_{r}^{R}$ and $\tilde\epsilon_{r}^{I}$ are the real and imaginary parts, respectively, of $\tilde{\epsilon}_r$, for $r = 1,\ldots,k$.  
 The limiting covariance matrix, $\tilde{\boldsymbol\Sigma}_{\tilde{\boldsymbol\xi}}$, only requires  the second moments of the predictor.
 Now, define the robust QDF as
 \begin{align}\label{eq:FT-RIRE}
    F_d (\boldsymbol{\Gamma}, \mathbf{C}; \tilde{\mathbf{G}}^{-1}) = [\text{vec}(\hat{\boldsymbol\xi}) - \text{vec}(\boldsymbol{\Gamma} \mathbf{C})]^T \tilde{\mathbf{G}}^{-1} [\text{vec}(\hat{\boldsymbol\xi}) - \text{vec}(\boldsymbol{\Gamma} \mathbf{C})],
 \end{align}
 where $ \tilde{\mathbf{G}} = (\mathbf{I} \otimes \hat{\boldsymbol\Sigma}^{-1/2}) \widehat{\text{cov}}[\text{vec}(\mathbf{Z}\tilde{\boldsymbol\epsilon}^T)] (\mathbf{I} \otimes \hat{\boldsymbol\Sigma}^{-1/2})$. The estimator that minimizes the robust QDF is called the Fourier transform robust inverse regression estimator (FT-RIRE). The inner product matrix $\mathbf{V} = \tilde{\mathbf{G}}^{-1}$ only needs second moments,  making FT-RIRE more theoretically robust.

 5. 
 Similarly, we define a diagonal block inner product matrix as $\tilde{\mathbf{G}}_D^{-1} = \text{diag}\{\tilde{\mathbf{G}}_1^{-1},\ldots,\tilde{\mathbf{G}}_s^{-1}\}$, where $\tilde{\mathbf{G}}_l^{-1}$ is defined for each $\boldsymbol \omega^{(l)} \in \R^{k_l}, l=1,\ldots, s$. The degenerated robust estimator that minimizes $F_d(\boldsymbol{\Gamma}, \mathbf{C};\tilde{\mathbf{G}}_D^{-1})$ is called the Fourier transform degenerated robust inverse regression estimator (FT-DRIRE).

\begin{algorithm}[H]
	\label{algo:5}
	\SetAlgoLined
	\begin{enumerate}
    \item Choose an initial value for $\boldsymbol{\Gamma}\in\mathbb{R}^{p\times d}$. One possible choices is to set $\boldsymbol e_i = (0,\cdots,0,1,0,\cdots,0)^T$ with the $i^{th}$ element equal to 1 and others equal to $0$. Alternatively, we use the Fourier transformation result from \cite{Weng18}.
    
    \item Fixed $\boldsymbol{\Gamma}$ and update $\mathbf{C}$ by minimizing $F_d(\boldsymbol{\Gamma},\mathbf{C};\mathbf{V})$. 
     Fit a linear regression of $\mathbf{V}^{1/2}\text{vec}(\hat{\boldsymbol\xi})$ on $\mathbf{V}^{1/2}(\mathbf{I}_{2m}\otimes \boldsymbol{\Gamma} )$, then obtain $\text{vec}(\mathbf{C}) = [(\mathbf{I}_{2m}\otimes \boldsymbol{\Gamma}^T) \mathbf{V} (\mathbf{I}_{2m}\otimes \boldsymbol{\Gamma} )]^{-1}(\mathbf{I}_{2m}\otimes \boldsymbol{\Gamma}^T)\mathbf{V} \text{vec}(\hat{\boldsymbol\xi})$.
    
    \item Fixed $\mathbf{C}$ and minimize $F_d(\boldsymbol{\Gamma}, \mathbf{C}; \mathbf{V})$ with respect to one column of $\boldsymbol{\Gamma}$, subject to the unit norm and orthogonality to other columns (keeping them constant). For this partial minimization problem, use the quadratic discrepancy function:  
    $F(\mathbf{b}) = (\boldsymbol\alpha_k - (\mathbf{c}_k^T \otimes \mathbf{I}_p) \mathbf{Q}_{\boldsymbol{\Gamma}_{(-k)}}\mathbf{b})^T \boldsymbol{V} (\boldsymbol\alpha_k - (\mathbf{c}_k^T \otimes \mathbf{I}_p) \mathbf{Q}_{\boldsymbol{\Gamma}_{(-k)}}\mathbf{b})
    $,
    where $\boldsymbol\alpha_k = \text{vec}(\hat{\boldsymbol\xi} - \boldsymbol{\Gamma}_{(-k)}\mathbf{C}_{(-k)})$, $\mathbf{c}_k$ is $k^{th}$ column of $\mathbf{C}$, $\mathbf{C}_{(-k)}$ (or $\boldsymbol{\Gamma}_{(-k)}$) is the matrix by deleting the $k^{th}$ column from $\mathbf{C}$ (or $\boldsymbol{\Gamma}$), and $\mathbf{Q}_{\boldsymbol{\Gamma}_{(-k)}}$ is orthogonal complement of Span$(\boldsymbol{\Gamma}_{(-k)})$. 
     Repeat the following steps for $k = 1,...,d$:
    \begin{enumerate}
    \item Let $\boldsymbol{\Gamma} = (\mathbf{b}_1,...,\mathbf{b}_{k-1},\mathbf{b}_k, \mathbf{b}_{k+1},...,\mathbf{b}_d)$ and update
    $
    \hat{\mathbf{b}}_k = \mathbf{Q}_{\boldsymbol{\Gamma}_{(-k)}}[\mathbf{Q}_{\boldsymbol{\Gamma}_{(-k)}}(\mathbf{c}_k^T \otimes \mathbf{I}_p) \mathbf{V}(\mathbf{c}_k^T \otimes \mathbf{I}_p) \mathbf{Q}_{\boldsymbol{\Gamma}_{(-k)}}]^{-} \mathbf{Q}_{\boldsymbol{\Gamma}_{(-k)}}(\mathbf{c}_k^T \otimes \mathbf{I}_p) \mathbf{V} \boldsymbol \alpha_k,
    $
    then normalize $\hat{\mathbf{b}}_k$ using $\hat{\mathbf{b}}_k / ||\hat{\mathbf{b}}_k||$.
    
    \item Update $\boldsymbol{\Gamma}$ by replacing $\mathbf{b}_k$ with $\hat{\mathbf{b}}_k$ and update $\mathbf{C}$ as described  in step 2. 
    \end{enumerate}
    \item Repeat   step 3 until the condition $\max\{|\boldsymbol{\Gamma}_{(t+1)}-\boldsymbol{\Gamma}_{(t)}|^2, |\mathbf{C}_{(t+1)}-\mathbf{C}_{(t)}|^2\} <10^{-6}$ is satisfied.
    \end{enumerate}
\caption{Minimum discrepancy approach}
\end{algorithm}

\subsection{R functions for the minimum discrepancy approaches}

Within the \textbf{itdr} package, the function  {\em fm\_xire()} provides estimators that utilize the minimum discrepancy approach with Fourier transformation. 
In this example, we use the   {\it prostate} dataset, which contains information o the level of a prostate-specific antigen associated with eight clinical measures in $97$ male patients who underwent a radical prostatectomy,  along with eight clinical measurements.  
 These clinical measurements are as follows: the logarithm of cancer volume (''lcavol``), the logarithm of prostate weight (''lweight``), age
(''age``), the logarithm of benign prostatic hyperplasia amount (''lbph), seminal vesicle invasion (''svi``), the logarithm of capsular penetration (''lcp``),
Gleason score (''gleason``), and the percentage of Gleason scores 4 or 5 (pgg45). The outcome variable is the logarithm of the prostate-specific antigen (''lpsa``). 
The following R code illustrates how to use the {\em fm\_xire()}    function.

\begin{example}
    library(itdr)
    set.seed(123)
    data(prostate)
    X=as.matrix(prostate[,1:8])
    Y=matrix(prostate[,9], ncol = 1)
    fit.ftire=fm_xire(Y,X,d=2,m = 10, method="FT-IRE")
    betahat = fit.ftire$hbeta_xire
    newx = X 
    plot(Y ~ newx[,1], xlab = "First reduced predictor", 
    ylab = paste0(expression(log),'(antigen)', sep="") )
    plot(Y ~ newx[,2], xlab = "Second reduced predictor", 
    ylab = paste0(expression(log),'(antigen)', sep="") )
\end{example}
\begin{singlespace}
\begin{example}
         [,1]          [,2]
[1,] -0.658371202 -0.0906078534
[2,] -0.611494712 -0.2493711238
[3,]  0.015269902 -0.0105628808
[4,] -0.148454529 -0.0753903192
[5,] -0.318437176  0.9142747970
[6,]  0.154861618  0.0360819508
[7,] -0.211976933 -0.2942920306
[8,] -0.005575356  0.0009348053
\end{example}   
\end{singlespace}

\begin{figure}
\centering
\subfigure{{\includegraphics[width=0.45\textwidth]{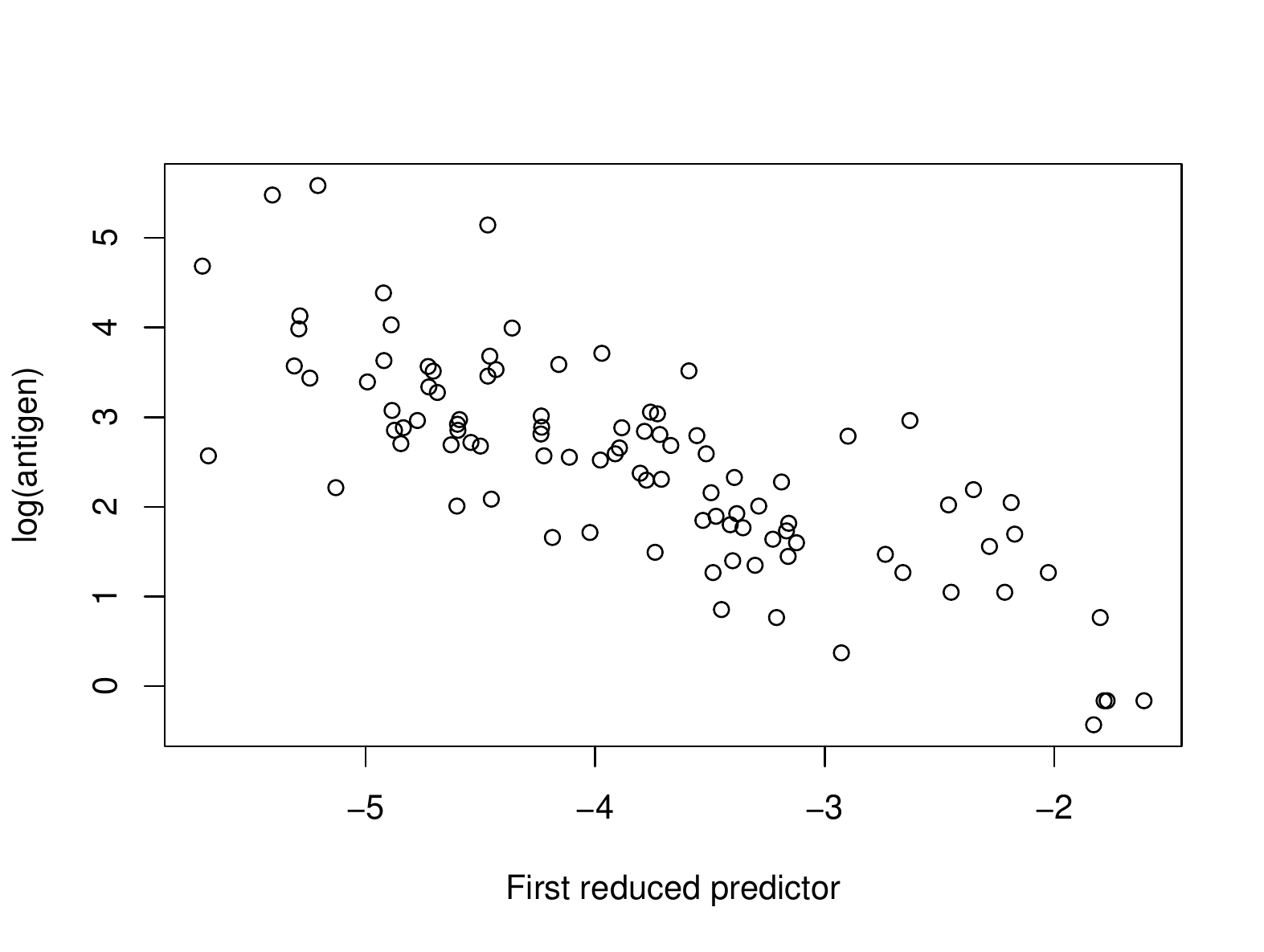} }}
\subfigure{{\includegraphics[width=0.45\textwidth]{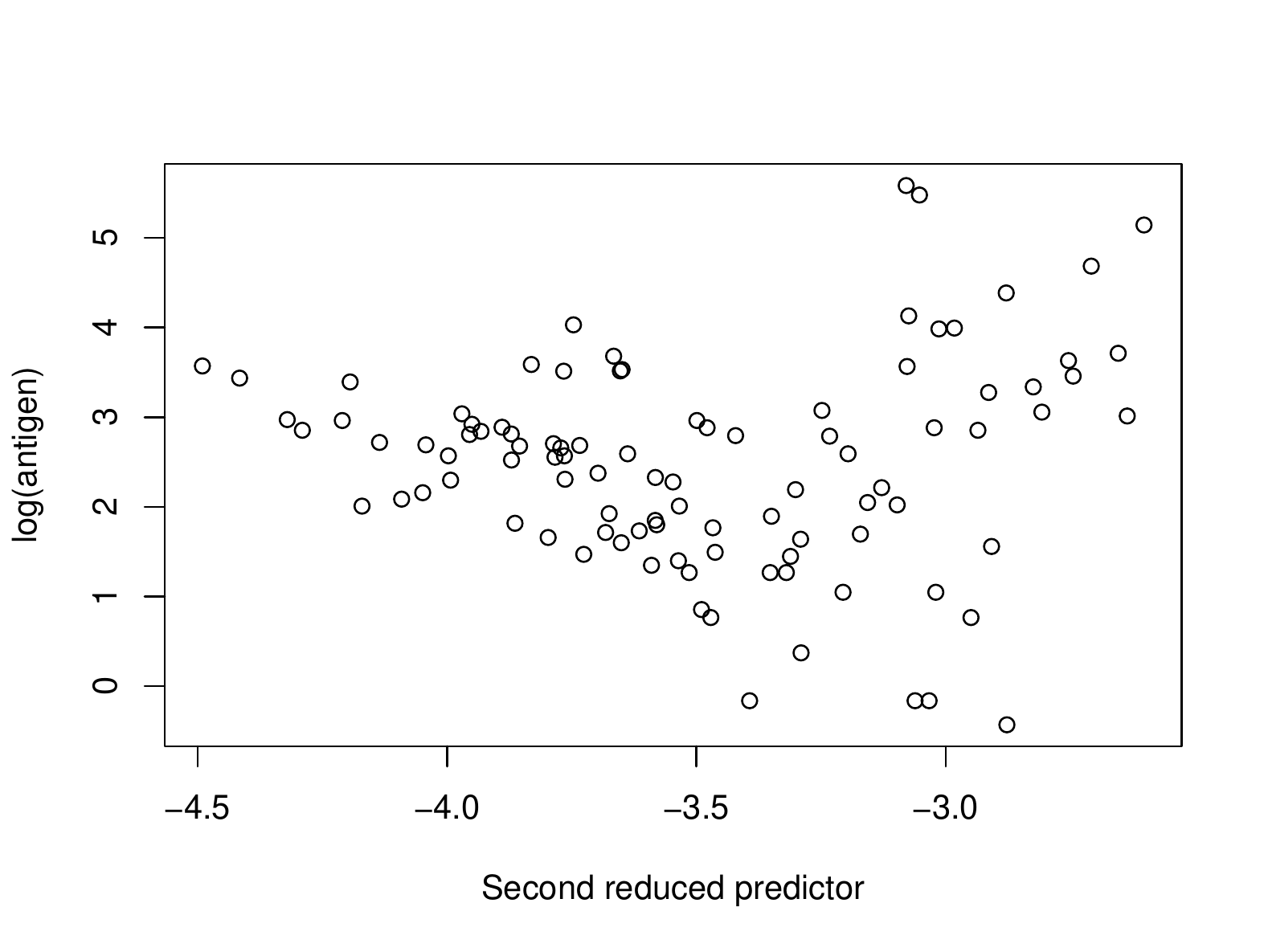} }}
\caption{The scatter plots of $\log$(antigen) versus the first two reduced predictors}
\label{fig:4}
\end{figure}

The {\em fm\_xire()} function includes the following arguments: {\em $\mathbf{Y}$}, representing the response matrix of dimension $n \times q$; {\em $\mathbf{X}$},  denoting the predictor matrix of dimension $n \times p$; {\em $d$}, specifying the dimension of the central subspace; {\em $m$}, indicating the number of Fourier transforms utilized in constructing the kernel matrix; and  {\em method}, which offers five options: FT-IRE, FT-DIRE, FT-SIRE, FT-RIRE, and FT-DRIRE, each corresponding to five different inner product matrix $\mathbf{V}$ in the QDF. 
As depicted in Figure \ref{fig:4}, the first two directions extracted by FT-IRE effectively capture both the linear and curvature relationships in the left and right panels, respectively.

\section{Fourier transform sparse inverse regression estimators for sufficient variable selection}

This section discusses the Fourier transform sparse inverse regression estimators proposed by \cite{weng2022fourier} in the high-dimensional regime, specifically when the sample size is smaller than the predictor dimension. 
It is often assumed that only a few predictors have a significant impact on the response variable. This implies   the existence of a  predictor subset, denoted by $S\in [p]$,  such that 
\begin{align*}
    \mathbf{Y} \independent \mathbf{X} \mid X_S.
\end{align*}
Equivalently, there exists a representation of $\boldsymbol{\Gamma}$ such that its nonzero rows are located  at the indices in the set $S$.
We consider a quadratic discrepancy function of $\boldsymbol{\Gamma} \in \mathbb{R}^{p\times d}$ and $\mathbf{C} \in \mathbb{R}^{d \times 2k}$,
\begin{equation}\label{eq:quad}
\begin{array}{cll} 
F(\boldsymbol{\Gamma}, \mathbf{C}) & = & [\text{vec}(\hat{\boldsymbol\xi}) - \text{vec}(\boldsymbol{\Gamma} \mathbf{C})]^T \text{diag}\{\hat{\boldsymbol\Sigma}\} [\text{vec}(\hat{\boldsymbol\xi}) - \text{vec}(\boldsymbol{\Gamma} \mathbf{C})] \\
& =  & \left\| \hat{\boldsymbol\Sigma}^{1/2} \hat{\boldsymbol\xi} - \hat{\boldsymbol\Sigma}^{1/2} \boldsymbol{\Gamma} C\right\|_F^2,
\end{array}
\end{equation}
where 
$\|\mathbf{A}\|_F$ is the Frobenius norm of matrix  $\mathbf{A}$ defined by $\sqrt{\text{trace}(\mathbf{A}\mathbf{A}^T)}$, and $\text{trace}(\cdot)$ denotes the trace of a matrix.
Sufficient variable selection aims to identify the indexes of active predictors $S = \{1 \leq j \leq p: \boldsymbol e_j^T \boldsymbol{\Gamma} \boldsymbol{\Gamma}^T \boldsymbol e_j > 0\}$. 
To achieve both sufficient dimension reduction and sufficient variable selection simultaneously, \cite{weng2022fourier} employed the coordinate-independent penalization \citep{chenzoucook10}, denoted as  $p_{\boldsymbol w}(\boldsymbol{\Gamma}) = \sum^p_{j=1} w_j \|\boldsymbol e_i^T \boldsymbol{\Gamma} \|_2$, where $\boldsymbol w = (w_1, \cdots, w_p)$ is the penalty weights. The objective function of interest is the quadratic function (\ref{eq:quad}) equipped with the coordinate-independent penalty $p_{\boldsymbol w}(\boldsymbol{\Gamma})$ and a tuning parameter $\lambda$. The optimization problem is as follows
\begin{equation}\label{eq:obj}
\begin{array}{ll}
\hat{\boldsymbol{\Gamma}} = & \arg\min_{\boldsymbol{\Gamma}, \mathbf{C}}\left\{ \frac{1}{2} \left\| \hat{\boldsymbol\Sigma}^{1/2} \hat{\boldsymbol\xi} - \hat{\boldsymbol\Sigma}^{1/2} \boldsymbol{\Gamma} \mathbf{C} \right\|_F^2 + \lambda p_{\boldsymbol w}(\boldsymbol{\Gamma}) \right\}, ~~~~ ~~~ \text{subject to}\ \mathbf{C}\mathbf{C}^T = \mathbf{I}_d.
\end{array}
\end{equation}
 Given $\mathbf{C}$, this problem  is convex with respect to  $\boldsymbol{\Gamma}$.
The nonsmooth penalty term $p_{\boldsymbol w}(\boldsymbol{\Gamma})$ encourages sparsity by shrinking small values of the rows in $\boldsymbol{\Gamma}$ towards zeros. The regularization parameter $\lambda$ controls model complexity.

\subsection{Algorithm for sparse estimators}\label{algo}


To solve the optimization problem (\ref{eq:obj}), \cite{weng2022fourier} considered an iterated alternating direction method of multipliers (ADMM) algorithm \citep{boyd2011distributed}. When solving for $\boldsymbol{\Gamma}$ given a  specific $\mathbf{C}$ value,   the following equivalent optimization problem is considered  
\begin{equation}\label{eq:algo1}
\begin{array}{cll}
&\arg\min_{\boldsymbol{\Gamma}}  &  \frac{1}{2} \left\| \hat{\boldsymbol\Sigma}^{1/2} \hat{\boldsymbol\xi} - \hat{\boldsymbol\Sigma}^{1/2} \boldsymbol{\Gamma} \mathbf{C} \right\|_F^2  + \sum_{j=1}^{p} \lambda_j \left\|  \mathbf{A}_{j\cdot}\right\|_2, ~~~~~~~~~~~~
 \mbox{subject to }  \boldsymbol{\Gamma} - \mathbf{A} = \mathbf{0},\\
\end{array}
\end{equation}
where $\lambda_j = \lambda w_j$ is the weight for each row of $\mathbf{A}$, i.e.,  $\mathbf{A}_{j\cdot}$. According to \cite{boyd2011distributed}, the augmented Lagrangian function over $\boldsymbol{\Gamma}$,  with the copy variable $\mathbf{A}$ and the scaled dual variable $\mathbf{U}$,  is defined as 
\begin{equation}\label{eq:algo2}
\begin{array}{cll}
 L_{\rho}(\boldsymbol{\Gamma}, \mathbf{A}, \mathbf{U}) & :=   &  \frac{1}{2} \left\| \hat{\boldsymbol\Sigma}^{1/2} \hat{\boldsymbol\xi} - \hat{\boldsymbol\Sigma}^{1/2} \boldsymbol{\Gamma} \mathbf{C} \right\|_F^2+ \sum_{j=1}^{p} \lambda_j \left\|  \mathbf{A}_{j\cdot}\right\|_2 + \frac{\rho}{2} \left\| \boldsymbol{\Gamma} - \mathbf{A} + \mathbf{U}\right\|_F^2,
\end{array}
\end{equation}
where $\rho > 0$ is the algorithm tuning parameter. Let $\boldsymbol\Xi = \boldsymbol\Sigma \boldsymbol\xi$ and $\hat{\boldsymbol\Xi} = \hat{\boldsymbol\Sigma} \hat{\boldsymbol\xi}$.  

To minimize Equation \eqref{eq:algo2} over $(\boldsymbol{\Gamma}, \mathbf{A}, \mathbf{U})$, we iterate the following three steps: 
\begin{align}
\boldsymbol{\Gamma}^{k+1} & = \argmin_{\boldsymbol{\Gamma}} L_{\rho}(\boldsymbol{\Gamma}, \mathbf{A}^k, \mathbf{U}^k), \label{eq:admmB}\\
\mathbf{A}^{k+1} & = \argmin_{\mathbf{A}} L_{\rho}(\boldsymbol{\Gamma}^{k+1}, \mathbf{A}, \mathbf{U}^k), \label{eq:admmA}\\
\mathbf{U}^{k+1} & = \mathbf{U}^{k} + \boldsymbol{\Gamma}^{k+1} - \mathbf{A}^{k+1}, \label{eq:admmU}
\end{align}
where $(\boldsymbol{\Gamma}^k, \mathbf{A}^k, \mathbf{U}^k)$ denotes the $k^{th}$ iteration. 
The explicit solutions corresponding to each step are stated in Algorithm~\ref{algo:6}.


\begin{algorithm}[H]
\label{algo:6}
\SetAlgoLined
\begin{enumerate}
    \item Update $\boldsymbol{\Gamma}^{k+1} = (\hat{\boldsymbol\Sigma} + \rho \mathbf{I}_p)^{-1}(\hat{\boldsymbol\Xi}\mathbf{C}^T + \rho \mathbf{A}^{k} - \rho \mathbf{U}^{k} )$.
    \item Update $\mathbf{A}^{k+1}_{j\cdot} = \max \left\{ 1 - \frac{\lambda_j/\rho}{ \left\| \boldsymbol{\Gamma}_{j\cdot}^{k+1} +  \mathbf{U}_{j\cdot}^{k} \right\|_2}, 0\right\} (\boldsymbol{\Gamma}_{j\cdot}^{k+1} +  \mathbf{U}_{j\cdot}^{k} )$ for $j = 1,\cdots, p, $ and $\lambda_j = \lambda w_j$.
    \item Update $\mathbf{U}^{k+1} = \mathbf{U}^{k} + \boldsymbol{\Gamma}^{k+1} - \mathbf{A}^{k+1}$.
    \item Repeat steps 1-3 until $\left\| \boldsymbol{\Gamma}^{k+1} - \mathbf{A}^{k+1} \right\|_F< \epsilon$ and $\left\|\rho (\mathbf{A}^{k} - \mathbf{A}^{k+1})\right\|_F < \epsilon.$
\end{enumerate}
\caption{Updating $\boldsymbol{\Gamma}$ given $\mathbf{C}$}
\end{algorithm}

Given $\boldsymbol{\Gamma}$ from Algorithm \ref{algo:6}, we further 
update $\mathbf{C}$ based on the following optimization problem because the regularization term does not involve $\mathbf{C}$: 
$$
\begin{array}{rl}
&  {\arg\min}_{\mathbf{C}} \|\hat{\boldsymbol\Sigma}^{1/2}\hat{\boldsymbol\xi} - \hat{\boldsymbol\Sigma}^{1/2} \boldsymbol{\Gamma} \mathbf{C} \|_F^2, ~~~~~~~~
\ \text{subject to}\ \boldsymbol{C}\mathbf{C}^T = \mathbf{I}_d.\\
\end{array} 
$$
The solution has a closed-form expression, that is $\hat{\mathbf{C}} = \mathbf{W}_2\mathbf{W}_1^T,$ where $\mathbf{W}_1\mathbf{D}\mathbf{W}^T_2$ is the singular value decomposition of $\hat{\boldsymbol\Xi}^T  \boldsymbol{\Gamma}$. Let $L(\boldsymbol{\Gamma}, \mathbf{C}) = \frac{1}{2} \left\| \hat{\boldsymbol\Sigma}^{1/2} \hat{\boldsymbol\xi} - \hat{\boldsymbol\Sigma}^{1/2} \boldsymbol{\Gamma} \mathbf{C} \right\|_F^2 + \lambda p_{\mathbf{w}}(\boldsymbol{\Gamma})$. 

\begin{algorithm}[H]
\label{algo:7}
\SetAlgoLined
\begin{enumerate}
    \item \textbf{Initialize} the algorithm with $\boldsymbol{\Gamma}^{0}$ and  equal weight $\boldsymbol w^{0}$. Start with $j = 0$.
    \item Update $\boldsymbol{C}^{j+1}=\boldsymbol{W}_2\boldsymbol{W}_1^T$, where $\boldsymbol{W}_1\boldsymbol{D}\boldsymbol{W}^T_2$ = SVD($\widehat{\boldsymbol\Xi}^T  \boldsymbol{\Gamma}^j$) and $\boldsymbol{\Gamma}^{j+1}$ using Algorithm \ref{algo:6} with input $\boldsymbol{C}^{j+1}$.
    \item Repeat step 2 with $j = j + 1$ or stop if $ |L(\boldsymbol{\Gamma}^{j+1} , \boldsymbol{C}^{j+1} ) - L(\boldsymbol{\Gamma}^{j} , \boldsymbol{C}^{j} )| < \epsilon.$
    \item Update weights $\mathbf{w}_i = \frac{1}{\left\|\mathbf{e}^T_i\boldsymbol{\Gamma}\right\|_2^{1/2}}$ for $ i = 1,\cdots,p.$ 
    \item Finally, repeat steps 2-3 using the new weights.
\end{enumerate}
\caption{Iterated ADMM}
\end{algorithm}

Algorithm \ref{algo:7} starts with equal weights $\mathbf{w}^0$, i.e., $(1,\cdots, 1)^T$, and then updates weights using $\boldsymbol{\Gamma}$ values. 
It is sufficient to update weights once to avoid overshrinkage of the estimation. 

\subsection{R functions for Fourier transform sparse inverse regression estimators}

In this section, we describe the {\em admmft()} function available in \textbf{itdr} package, which enables the selection of active variables using the Fourier transformation method \citep{weng2022fourier}. The following R codes demonstrate the application of this function for sufficient variable selection on the {\it Raman} dataset in \textbf{itdr} package. By default,  the tuning parameter $\lambda$ is set to $0.5$. 
However, if no specific value is provided, the function utilizes cross-validation to determine the optimal {\em lambda} value.    

\begin{example}
    data(raman)
    Y=as.matrix(Raman[,c(1100)]) ## percentage of total fat content
    X=as.matrix(Raman[c(2:501)]) ## first 500 wavelength variables
    out = admmft(X,Y,d = 1, m = 30, lambda = 0.5, sparse.cov=T, scale.X=T)
    estbeta = out$B
    plot(Y ~ X 
    ylab = "Percentage of total fat")
\end{example}
The {\em admmft()} function accepts the following arguments: 
{\em $\mathbf{X}$}, the predictor matrix of dimension $n \times p$; {\em $\mathbf{Y}$}, the response matrix of dimension $n \times q$; {\em $d$}, the dimension of the central subspace; 
{\em $m$}, the number of Fourier transforms used in constructing the kernel matrix; 
{\em lambda}, the tuning parameter. If it is not provided, then the optimal lambda value is chosen by cross-validation using the Fourier transformation method;
{\em noB}, the number of iterations for updating B, the default value is $5$;
{\em noC}, the number of iterations for updating C, the default value is $20$;
{\em noW}, the number of iterations for updating the weight, the default value is $2$;
{\em sparse.cov},  a logical value that determines whether to calculate the soft-threshold matrix for the covariance matrix. If set to TRUE, the soft-threshold matrix is computed.
{\em  scale.X}, a logical value that determines whether to standardize each variable when calculating the soft-threshold matrix for the covariance matrix. If set to TRUE, variables are standardized. 

\begin{figure}
\centering
{\includegraphics[width=0.5\textwidth]{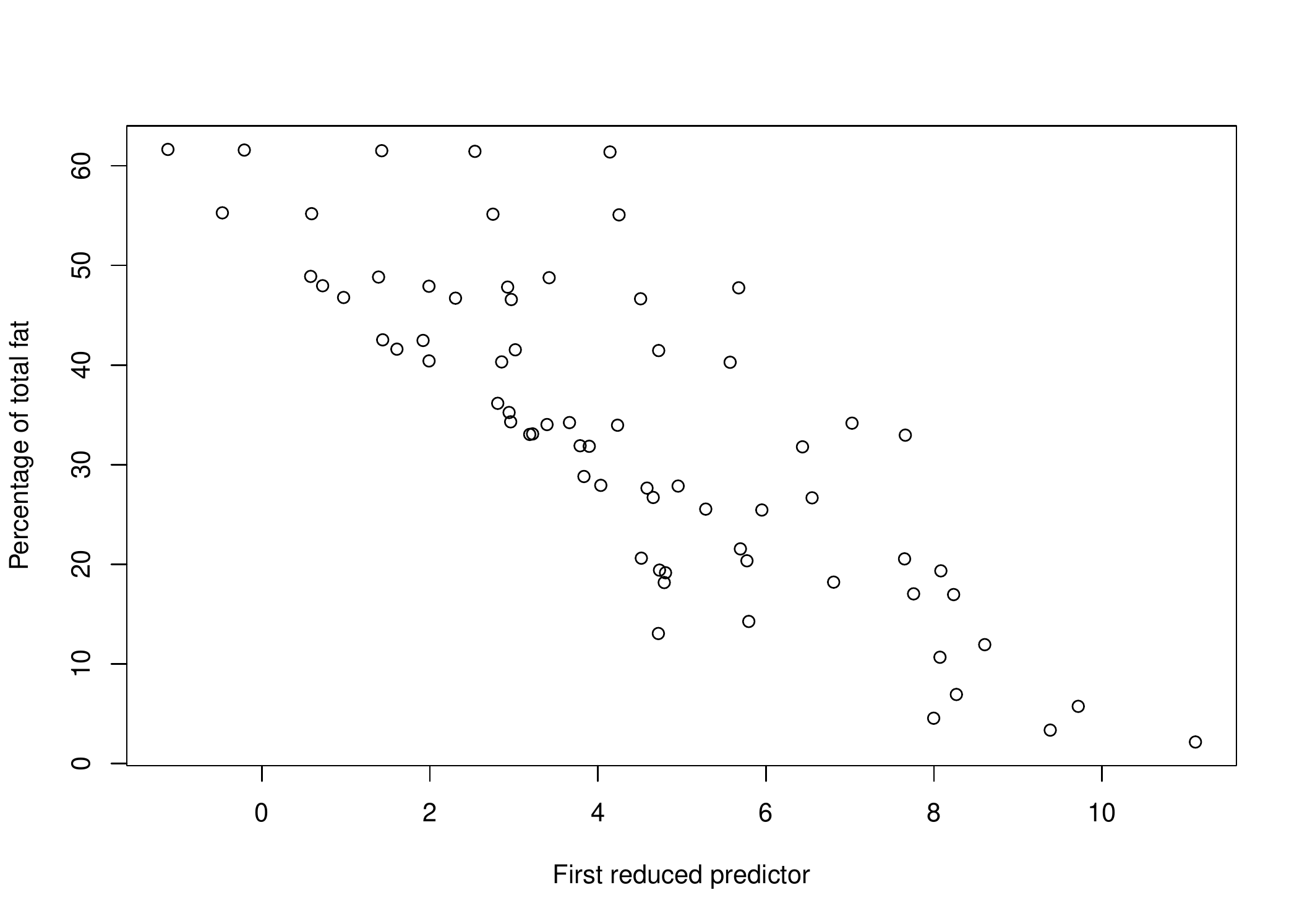} }
\caption{The scatter plot of $\log$(antigen) versus the first two reduced predictors}
\label{fig:5}
\end{figure}

Based on the results shown in Figure \ref{fig:5}, it can be inferred that the first direction obtained from the ADMM Fourier transformation approach exhibits a discernible downward trend in relation to the percentage of total fat content.

\section{Summary}
This paper has introduced the $\mathbf{itdr}$ R package   which offers a comprehensive set of functions for estimating the central subspace (CS) and the central mean subspace (CMS) using integral transformation methods. 
We have provided an overview of the sufficient dimension reduction technique and discussed various integral transformation methods, including the Fourier transformation method, the convolution transformation method, the iterative Hessian transformation method, the Fourier transformation approach for inverse regression, the minimum discrepancy approach, and Fourier transform sparse inverse regression. 
The first three methods are specifically designed for univariate responses, whereas the latter three methods are applicable to both univariate and multivariate responses.
The $\mathbf{itdr}$ package equips users with powerful tools to estimate the dimension and sufficient dimension reduction subspaces. Additionally, it provides essential functions and options to relax the normality assumption through the estimation of the density function using the kernel smoothing method.  These features expand the potential applications of the package to a wider range of domains.  


\bibliography{Tharindu_DeAlwis}

\begin{thebibliography}{30}
\providecommand{\natexlab}[1]{#1}
\providecommand{\url}[1]{\texttt{#1}}
\expandafter\ifx\csname urlstyle\endcsname\relax
  \providecommand{\doi}[1]{doi: #1}\else
  \providecommand{\doi}{doi: \begingroup \urlstyle{rm}\Url}\fi

\bibitem[Adragni and Raim(2014)]{ldr2014}
K.~P. Adragni and A.~M. Raim.
\newblock ldr: An r software package for likelihood-based sufficient dimension
  reduction.
\newblock \emph{Journal of Statistical Software}, 61:\penalty0 1--21, 2014.

\bibitem[Bentler and Xie(2000)]{Bentler2000}
P.~M. Bentler and J.~Xie.
\newblock Corrections to test statistics in principal hessian directions.
\newblock \emph{Statistics and Probability Letters}, 47:\penalty0 381--389,
  2000.

\bibitem[Boyd et~al.(2011)Boyd, Parikh, Chu, Peleato, Eckstein,
  et~al.]{boyd2011distributed}
S.~Boyd, N.~Parikh, E.~Chu, B.~Peleato, J.~Eckstein, et~al.
\newblock Distributed optimization and statistical learning via the alternating
  direction method of multipliers.
\newblock \emph{Foundations and Trends{\textregistered} in Machine learning},
  3\penalty0 (1):\penalty0 1--122, 2011.

\bibitem[Chen et~al.(2010)Chen, Zou, and Cook]{chenzoucook10}
X.~Chen, C.~Zou, and R.~D. Cook.
\newblock Coordinate-independent sparse sufficient dimension reduction and
  variable selection.
\newblock \emph{Annals of Statistics}, 38\penalty0 (6):\penalty0 3696--3723,
  2010.

\bibitem[Clark et~al.(1987)Clark, Henderson, Hoggard, Ellison, and
  Young]{Clark87}
R.~G. Clark, H.~V. Henderson, G.~K. Hoggard, R.~S. Ellison, and B.~J. Young.
\newblock The ability of biochemical and haematological tests to predict
  recovery in periparturient recumbent cows.
\newblock \emph{NZ Veterinary Journal}, 35:\penalty0 126--133, 1987.

\bibitem[Cook(1998)]{cook98}
R.~D. Cook.
\newblock \emph{Regression Graphics: Ideas for Studying Regressions Through
  Graphics}.
\newblock New York: Wiley, 1998.

\bibitem[Cook(2007)]{cook07}
R.~D. Cook.
\newblock {Fisher Lecture: Dimension Reduction in Regression}.
\newblock \emph{Statistical Science}, 22\penalty0 (1):\penalty0 1 -- 26, 2007.

\bibitem[Cook and Forzani(2008a)]{cookfroz08a}
R.~D. Cook and L.~Forzani.
\newblock Covariance reducing models: An alternative to spectral modeling of
  covariance matrices.
\newblock \emph{Biometrika}, 95\penalty0 (4):\penalty0 799--812, 2008a.

\bibitem[Cook and Forzani(2008b)]{cookfro08b}
R.~D. Cook and L.~Forzani.
\newblock Principal fitted components for dimension reduction in regression.
\newblock \emph{Statistical Science}, 23\penalty0 (4):\penalty0 485--501,
  2008b.

\bibitem[Cook and Forzani(2009)]{lbase}
R.~D. Cook and L.~Forzani.
\newblock Likelihood-based sufficient dimension reduction.
\newblock \emph{Journal of the American Statistical Association}, 104:\penalty0
  197--208, 2009.

\bibitem[Cook and Li(2002)]{CL02}
R.~D. Cook and B.~Li.
\newblock Dimension reduction for the conditional mean in regression.
\newblock \emph{Annals of Statistics}, 30:\penalty0 455--474, 2002.

\bibitem[Cook and Ni(2005)]{CN05}
R.~D. Cook and L.~Ni.
\newblock Sufficient dimension reduction via inverse regression: A minimum
  discrepancy approach.
\newblock \emph{Journal of the American Statistical Association}, 100:\penalty0
  410--428, 2005.

\bibitem[Cook and Weisberg(1991)]{CW91}
R.~D. Cook and S.~Weisberg.
\newblock Sliced inverse regression for dimension reduction: Comment.
\newblock \emph{Journal of the American Statistical Association}, 86:\penalty0
  328--332, 1991.

\bibitem[Feng et~al.(2013)Feng, Wen, Yu, and Zhu]{psave2013}
Z.~Feng, M.~X. Wen, Z.~Yu, and L.~Zhu.
\newblock On partial sufficient dimension reduction with applications to
  partially linear multi-index models.
\newblock \emph{Journal of the American Statistical Association}, 108:\penalty0
  236--246, 2013.

\bibitem[Folland(1992)]{Folland92}
G.~B. Folland.
\newblock \emph{Fourier Analysis and its Applications}.
\newblock Brooks/Cole, 1992.

\bibitem[Hang and Xia(2019)]{Weiq2019}
W.~Hang and Y.~Xia.
\newblock \emph{MAVE: Methods for Dimension Reduction}.
\newblock The Comprehensive R Archive Network, 2019.
\newblock URL \url{https://CRAN.R-project.org/package=MAVE}.

\bibitem[Hristache et~al.(2001)Hristache, Juditsky, Polzehl, and Spokoiny]{H01}
M.~Hristache, A.~Juditsky, J.~Polzehl, and V.~G. Spokoiny.
\newblock Structure adaptive approach for dimension reduction.
\newblock \emph{Annals of Statistics}, 29:\penalty0 1537--1566, 2001.

\bibitem[Li(2018)]{lib2018}
B.~Li.
\newblock \emph{Sufficient dimension reduction: Methods and applications with
  R}.
\newblock CRC Press, Boca Raton, FL, 2018.

\bibitem[Li and Kim(2021)]{nsdr2021}
B.~Li and K.~Kim.
\newblock \emph{nsdr: Nonlinear Sufficient Dimension Reduction}.
\newblock The Comprehensive R Archive Network, 2021.
\newblock URL \url{https://CRAN.R-project.org/package=nsdr}.

\bibitem[Li(1991)]{Li91}
K.~C. Li.
\newblock Sliced inverse regression for dimension reduction.
\newblock \emph{Journal of the American Statistical Association}, 86\penalty0
  (414):\penalty0 316--327, 1991.

\bibitem[Li(1992)]{Li92}
K.~C. Li.
\newblock On principal hessian directions for data visualization and dimension
  reduction: Another application of stein's lemma.
\newblock \emph{Journal of the American Statistical Association}, 87:\penalty0
  1025--1039, 1992.

\bibitem[Ma and Zhu(2013)]{eff2013}
Y.~Ma and L.~Zhu.
\newblock Efficient estimation in sufficient dimension reduction.
\newblock \emph{Annals of Statistics}, 41:\penalty0 250--268, 2013.

\bibitem[Weisberg(2015)]{Weis2015}
S.~Weisberg.
\newblock \emph{dr: Methods for Dimension Reduction for Regression}.
\newblock The Comprehensive R Archive Network, 2015.
\newblock URL \url{https://CRAN.R-project.org/package=dr}.

\bibitem[Weng(2022)]{weng2022fourier}
J.~Weng.
\newblock Fourier transform sparse inverse regression estimators for sufficient
  variable selection.
\newblock \emph{Computational Statistics \& Data Analysis}, 168:\penalty0
  107380, 2022.

\bibitem[Weng and Yin(2018)]{Weng18}
J.~Weng and X.~Yin.
\newblock Fourier transform approach for inverse dimension reduction method.
\newblock \emph{Journal of Nonparametric Statistics}, 30\penalty0 (4):\penalty0
  1049--1071, 2018.

\bibitem[Weng and Yin(2022)]{weng2022minimum}
J.~Weng and X.~Yin.
\newblock A minimum discrepancy approach with fourier transform in sufficient
  dimension reduction.
\newblock \emph{Statistica Sinica}, 32:\penalty0 2381--2403, 2022.

\bibitem[Xia et~al.(2002)Xia, Tong, Li, and Zhu]{X02}
Y.~Xia, H.~Tong, W.~Li, and L.~X. Zhu.
\newblock An adaptive estimation of dimension reduction.
\newblock \emph{Journal of the Royal Statistical Society. Series B},
  64:\penalty0 363--410, 2002.

\bibitem[Zeng and Zhu(2010)]{zeng2010}
P.~Zeng and Y.~Zhu.
\newblock An integral transform method for estimating the central mean and
  central subspaces.
\newblock \emph{Journal of Multivariate Analysis}, 101\penalty0 (1):\penalty0
  271--290, 2010.

\bibitem[Zhu et~al.(2019)Zhu, Zhang, Zhao, PengXu, Zhou, and
  XinZhang]{orthodr2019}
R.~Zhu, J.~Zhang, R.~Zhao, PengXu, W.~Zhou, and XinZhang.
\newblock orthodr: Semiparametric dimension reduction via orthogonality
  constrained optimization.
\newblock \emph{The R Journal}, 11:\penalty0 24--37, 2019.

\bibitem[Zhu and Zeng(2006)]{zhu2006}
Y.~Zhu and P.~Zeng.
\newblock Fourier methods for estimating the central subspace and the central
  mean subspace in regression.
\newblock \emph{Journal of the American Statistical Association}, 101:\penalty0
  1638--1651, 2006.

\end{thebibliography}

\address{Tharindu P. De Alwis\\
  School of Mathematical and Statistical Sciences,\\
  Southern Illinois University Carbondale\\
  1245 Lincoln Drive,\\
  Carbodnale, IL-62901\\
  United States\\
  (0000-0002-3446-0502)\\
  \email{mktharindu87@siu.edu}}

\address{S. Yaser Samadi\\
  School of Mathematical and Statistical Sciences,\\
  Southern Illisnois University Carbondale\\
   1245 Lincoln Drive,\\
 Carbodnale, IL-62901\\
 United States\\
  (0000-0002-6121-0234)\\
  \email{ysamadi@siu.edu}}

\address{Jiaying Weng\\
  Department of Mathematical Sciences,\\
  Bentley University\\
   175 Forest Street,\\
 Waltham, MA-02452\\
 United States\\
  (0000-0002-9463-5714)\\
  \email{jweng@bentley.edu}}
\end{article}

\end{document}